\documentclass[%
reprint,superscriptaddress,
amsmath,amssymb,
aps,
apl
]{revtex4-2}

\usepackage{graphicx}% Include figure files
\usepackage{dcolumn}% Align table columns on decimal point
\usepackage{bm}% bold math
\usepackage{float}
\usepackage{hyperref}
\hypersetup{hidelinks}
\hypersetup{colorlinks = true, linkcolor=blue, urlcolor=blue,citecolor=blue}
\usepackage{physics}
\usepackage{suffix}
\usepackage{xstring}
\usepackage[usenames, dvipsnames]{color}
\usepackage{gitinfo}
\usepackage{siunitx}
\DeclareSIUnit\bohr{bohr}
\usepackage[backgroundcolor=blue!20!white]{todonotes}
\usepackage[normalem]{ulem}

\def \br{{\bf r}}

\def \mum{\mu \mathrm{m}}

\def \m2D{\mathrm{2D}}

\def \mms{\mathrm{ms}}

\def \mHz{\mathrm{Hz}}
\newcommand*{\Rb}[1]{\ensuremath{\mathrm{^{#1}Rb}}}

\begin{document}
	
	\preprint{APS/123-QED}
	
	%\title{Dynamical BKT scaling in a quenched 2D Bose gas}
	\title{Universal Scaling of the Dynamic BKT Transition in Quenched 2D Bose Gases}
	%\title{Universal dynamics of the BKT transition in quenched 2D Bose gases}
	%\title{Universal scaling of the critical dynamics in quenched 2D Bose gases}

	\author{S. Sunami}%
	\email{shinichi.sunami@physics.ox.ac.uk}
	\affiliation{Clarendon Laboratory, University of Oxford, Oxford OX1 3PU, United Kingdom}
	
	\author{V. P. Singh}
	\affiliation{Institut f\"ur Theoretische Physik, Leibniz Universit\"at Hannover, Appelstra{\ss}e 2, 30167 Hannover, Germany}
	\affiliation{Zentrum f\"ur Optische Quantentechnologien and Institut f\"ur Laserphysik, Universit\"at Hamburg, 22761 Hamburg, Germany}
	
	\author{D. Garrick}%
	\affiliation{Clarendon Laboratory, University of Oxford, Oxford OX1 3PU, United Kingdom}
	
	\author{A. Beregi}%
	\affiliation{Clarendon Laboratory, University of Oxford, Oxford OX1 3PU, United Kingdom}

	\author{A. J. Barker}%
	\affiliation{Clarendon Laboratory, University of Oxford, Oxford OX1 3PU, United Kingdom}
	
	\author{\\K. Luksch}%
	\affiliation{Clarendon Laboratory, University of Oxford, Oxford OX1 3PU, United Kingdom}
	
	\author{E. Bentine}%
	\affiliation{Clarendon Laboratory, University of Oxford, Oxford OX1 3PU, United Kingdom}
	
	\author{L. Mathey}
	\affiliation{Zentrum f\"ur Optische Quantentechnologien and Institut f\"ur Laserphysik, Universit\"at Hamburg, 22761 Hamburg, Germany}
	\affiliation{The Hamburg Centre for Ultrafast Imaging, Luruper Chaussee 149, Hamburg 22761, Germany}
	
	\author{C. J. Foot}%
	\affiliation{Clarendon Laboratory, University of Oxford, Oxford OX1 3PU, United Kingdom}

	\date{\today}
	
	\begin{abstract}
		\noindent 
		While renormalization group theory is a fully established method to capture equilibrium phase transitions, the applicability of RG theory to universal non-equilibrium behavior remains elusive.
		Here we address this question by measuring the non-equilibrium dynamics triggered by a quench from superfluid to thermal phase across the Berezinskii-Kosterlitz-Thouless transition in a 2D Bose gas.
		We quench the system by splitting the 2D gas in two and probe the relaxation dynamics by measuring the phase correlation function and vortex density via matter-wave interferometry.
		The dynamics occur via a two-step process of rapid phonon thermalization followed by slow dynamic vortex unbinding. 
		We demonstrate universal scaling laws for the algebraic exponents and vortex density, supported by classical-field simulations, and show their agreement with the real-time RG theory.
	\end{abstract}
	
	\maketitle	
	The relaxation dynamics of a many-body system that is quenched out of equilibrium displays a wide range of scenarios, from simple exponential decay to relaxation via metastable or prethermalized states \cite{Langen2015,Schweigler2017}, including phenomena such as pattern formation \cite{Zahn2022}, and the absence of thermalization \cite{Kinoshita2006}.   
	Systems that are quenched across a phase transition are particularly intriguing because of their universal self-similar behavior, expected in systems as diverse as superfluid helium \cite{Zurek1985}, liquid crystals \cite{Chuang1991}, biological cell membranes \cite{Sarah2007}, the early universe \cite{Kibble1976}, and cold atoms \cite{Zhou2010, Polkovnikov2011}. 
	There are numerous theoretical challenges in the treatments of non-equilibrium dynamics, see e.g. \cite{Eisert2015}, and this motivates in-depth experimental studies to guide and test theories.
	
	%A platform of unprecedented control and tunability for this purpose are ultracold gases,
	For this purpose, ultracold gases have emerged as a platform of unprecedented control and tunability,  which serve as quantum simulators for the investigation of many-body dynamics. 
	This has led to the observation of Kibble-Zurek (KZ) scaling \cite{Navon2015, Clark2016, Beugnon2017, Keesling2019} and universal scaling laws  \cite{Prufer2018,Erne2018, Glidden2021} following a quench. 
	These cold-atom experiments, however, mostly measure global observables, except for special cases such as the local probe of 1D spinor gas \cite{Prufer2018}.
	%This motivates to apply the technique of local matter-wave interferometry, previously utilised to probe the local phase fluctuation of near-integrable 1D systems \cite{Langen2015,Schweigler2017}, to investigate microscopic feature of critical phenomena in higher dimensions.
	%To understand the microscopic origin of universal dynamics, a promising method is the direct probe of fluctuation through the extension of local matter-wave interferometry, previously utilised to probe the local phase fluctuation of near-integrable 1D systems \cite{Langen2015,Schweigler2017}, to investigation of critical phenomena in higher dimensions \cite{Sunami2021}.
	To understand the microscopic origin of universal dynamics, a promising method is to directly probe fluctuations through the extension of local matter-wave interferometry, previously utilised to probe the local phase fluctuations of near-integrable 1D systems \cite{Langen2015,Schweigler2017}, to the investigation of critical phenomena observable in higher dimensions \cite{Sunami2021}.
	
	In 2D an especially interesting case is the critical dynamics across the Berezinskii-Kosterlitz-Thouless (BKT) transition  \cite{Berezinskii1972,Kosterlitz1973,Nelson1977} when the system is quenched from the superfluid to the thermal phase. 
	Real-time renormalization-group (RG) theory and truncated Wigner simulations \cite{Mathey2010} predict that the relaxation occurs via a reverse-Kibble-Zurek type mechanism, in which delayed vortex proliferation results in a metastable supercritical phase.
	The BKT transition is driven by the unbinding of vortex-antivortex pairs \cite{Hadzibabic2006, Sunami2021}, underscoring the topological nature of the transition.
	This transition is characterized by a sudden change of the functional form of the correlation function from power-law in the superfluid regime, $g_1(r,r')=\langle \Psi(r)^{\dagger}\Psi(r') \rangle \propto |r-r'|^{-\eta}$, to exponential in the thermal regime $g_1(r,r')  \propto e^{-|r-r'|/\xi}$,  where $\Psi(r)$ is the bosonic field operator at location $r$ and $\xi$ is the correlation length.
	The algebraic exponent $\eta$ has a universal value at the equilibrium critical point of $\eta_{\mathrm{BKT}}=0.25$ in the thermodynamic limit, however, the critical value of $\eta$ is strongly affected by finite-size effects \cite{Sunami2021}.
	
	Here, we study the critical dynamics across the BKT transition by quenching a 2D Bose gas from the superfluid to the thermal phase by splitting it in two. 
	Using spatially-resolved matter-wave interferometry, we measure the first-order correlation function and vortex density to analyze their relaxation dynamics. 
	We find that relaxation occurs via a two-step process, involving phonon relaxation and then dynamical vortex proliferation.
	We demonstrate universal scaling laws for the algebraic exponent and vortex density by performing measurements at different initial conditions.  
	Both real-time RG theory \cite{Mathey2010,Mathey2017} and classical-field  simulations are in good agreement with our measurements.

	\begin{figure*}
		\centering
		\includegraphics[width=0.8 \textwidth]{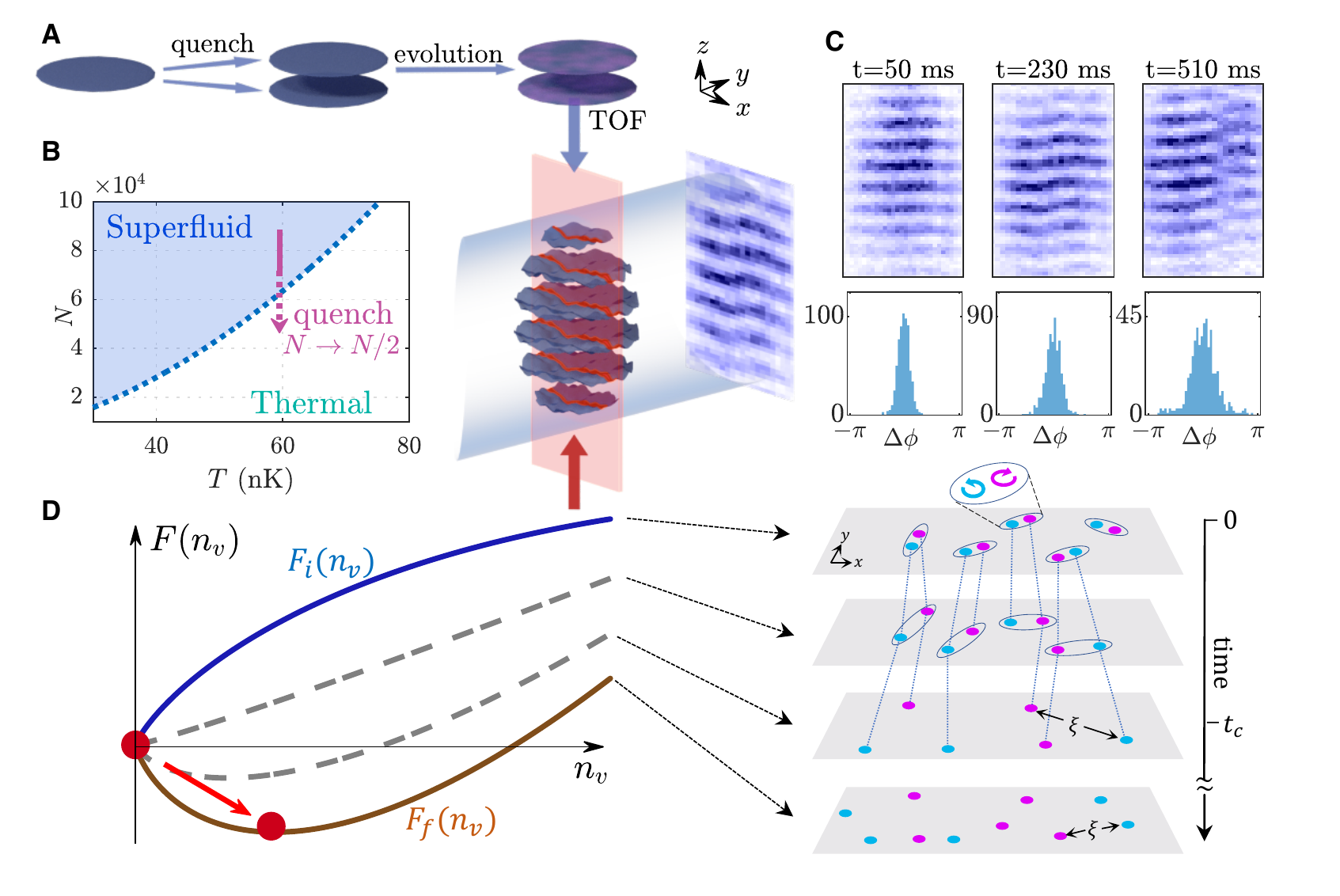}
		\caption{\textbf{Observation of non-equilibrium dynamics in 2D Bose gases via matter-wave interferometry.}
			(\textbf{A}) A 2D superfluid is split into two daughter clouds, thereby quenching through the BKT transition.
			The two clouds evolve for time $t$ and are released to produce matter-wave interference after a time-of-flight (TOF).
			Local phase fluctuations are observed by optically pumping the slice (red sheet) and then performing absorption imaging.
			(\textbf{B}) Equilibrium phase-diagram of trapped 2D Bose gases \cite{supp}. 
			The quench forces the system out of equilibrium towards the thermal phase. 
			(\textbf{C}) Examples of interference images (top).
			Phase dislocation due to the presence of a vortex is visible in the image at 510ms.
			The histograms (bottom) show the phase differences $\Delta \phi = \phi(x)-\phi(x')$ at $|x-x'|=5 \mu$m from 45 experimental runs.
			The decreasing height and increasing width indicate increased phase fluctuations. 
			(\textbf{D}) (left) Free energies $F_i(n_v)$ and $F_{f}(n_v)$ (continuous lines) in equilibrium, for the initial and final conditions of the quench with their minimum values indicated by red points \cite{Wen2010}. 
			Following the quench, the system relaxes towards the state with nonzero free vortex density $n_v$.
			(right) Illustration of the dynamics showing the transition between a scale-invariant phase supported by bound vortex-antivortex pairs and the broken scale-invariant phase characterized by free vortices with the mean vortex-vortex distance determining the correlation length $\xi$ where $t_c$ is the crossover time. 
		}
		\label{fig:main} 
	\end{figure*}	
	%\newpage
	
	Our experiments begin with a single pancake-shaped quasi-2D Bose gas in the superfluid regime, consisting of $N \approx 9\times 10^4$ atoms of $^{87}$Rb at reduced temperatures in the range $\tilde{T}= 0.3-0.5$, where $\tilde{T} \equiv T/T_0$ is the ratio of the initial temperature $T$ and the critical temperature $T_0$ of a non-interacting trapped 	gas \cite{T0}.
	The quench is implemented by a rapid splitting of the system in a multiple-RF dressed potential \cite{supp,Barker2020,Harte2018,Barker2020jphysb}, as illustrated in Fig.\ \ref{fig:main}A,  which results in a pair of decoupled clouds each with atom number $N'=N/2$ trapped in the two minima of a double-well potential.
	Each well has vertical trap frequencies of $\omega_z/2\pi  = 1$ kHz and the dimensionless 2D interaction strength is $\tilde{g} = 0.076$ \cite{supp}.
	The initial $\tilde{T}$ is chosen so that the value after splitting corresponds to the thermal phase if the system was in equilibrium (Fig.\ \ref{fig:main}B). 
	To investigate the dynamics, we let each cloud evolve independently for time $t$ before performing a time-of-flight expansion of $t_\mathrm{TOF}=16$ ms after which we detect the matter-wave interference that encodes the \textit{in situ} relative phase fluctuation of two clouds $ \phi (x)$ along a line that goes through the center of the cloud.
	%, which we use to characterize the non-equilibrium dynamics. 
	Interference images and histograms of spatial phase fluctuations $\Delta \phi $ show stronger fluctuations at long evolution times (Fig.\ 1C). 
	The dynamics across the BKT transition is expected to be scale-invariant until the bound vortex-antivortex pairs dissociate to disrupt the phase coherence (Fig.\ 1D).
	
	To analyze the relaxation dynamics quantitatively, we use the interference pattern to determine both the correlation function and the vortex density \cite{Sunami2021}. 
	The correlation function is defined as $C(r) = \mathrm{Re} \Big[\langle e^{i\phi(x-r/2)-i\phi(x+r/2)} \rangle_{x \in w} \Big]$, 
	where $\langle .. \rangle$ denotes an ensemble average of $N_r=45$ experimental repeats as well as average within the region of interest $w$ such that we perform the analysis only where a clear interference pattern is observed \cite{supp}.
	At each evolution time $t$, the degree of correlation of the phases at points separated by distance $r$ is quantified by $C(r)$, related to the first-order correlation function $g_1(r)$ by $C(r)\simeq g_1^2(r)/n^2$ in the absence of coupling between the two clouds, where $n$ is the 2D density \cite{supp,Mathey2017}.
	
	\begin{figure*}
		\includegraphics[width=0.87	\textwidth]{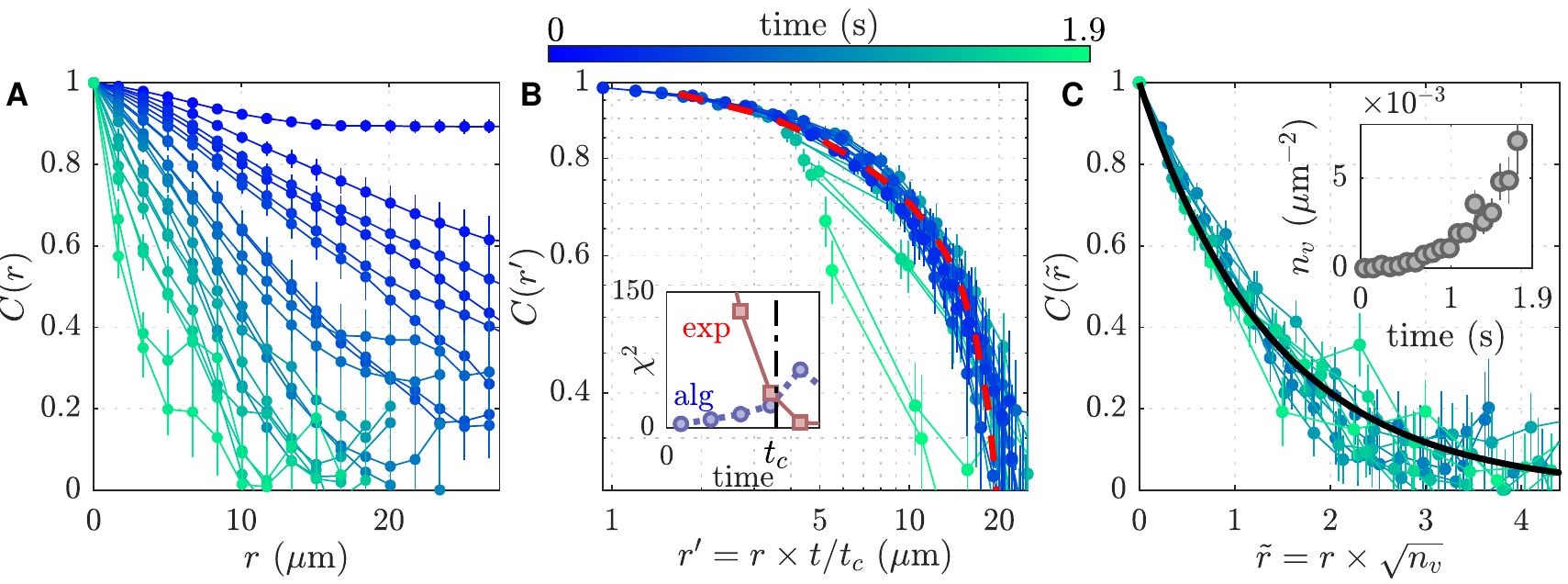}
		\caption{ \textbf{Non-equilibrium correlation functions and their scaling dynamics.} 
			(\textbf{A}) Relaxation dynamics of the phase correlation function $C(r)$ measured after the quench for the initial temperature $\tilde{T}=0.41$, where $C(r)$ is averaged over more than $45$ realizations and error bars denote standard error.
			(\textbf{B}) Linear scaling of length $r^\prime = r t/t_c$ according to time $t/t_c$ results in a collapse towards a common curve for time evolution up to $t \sim 1$ s. 
			This universal function is compatible with the expected power-law behavior at the crossover in equilibrium (red dashed line), including the effect of inhomogeneity \cite{supp}. 
			At long times, deviations (green curves) are observed, indicating the breaking of scale invariance. 
			Inset shows the $\chi^2$ errors of the algebraic and exponential fit functions \cite{supp}, which are used to determine the crossover time $t_c \sim 0.5 $ s. 
			(\textbf{C}) Temporal scaling of the distance $\tilde{r}=r\sqrt{n_v (t)}$  according to the vortex density $n_v(t)$ results in a collapse for times $t > 1$ s.
			Scaling behavior is compatible with an exponential decay (black continuous line).
			Inset shows $n_v(t)$ after the quench.
		}
		\label{fig:corrfunc}
	\end{figure*}	
	
	In Fig.\ \ref{fig:corrfunc}A, we show the time evolution of $C(r)$ after the quench at $t=0$. 
	Initially, there is almost no correlation decay because the two clouds have nearly identical phases;
	their phases decouple after a few tens of milliseconds. 
	After this, $C(r)$ begins to fall off for all $r$, and this fall-off increases as $t$ increases.
	At longer times, the correlation function drops sharply to 0, corresponding to no coherence at large distances. 
	This qualitative change of $C(r)$ indicates a dynamical transition, where the system relaxes to the high-temperature phase. 
	To determine the nature of the transition, we fit $C(r)$ with the algebraic and exponential functions which are used to characterize the equilibrium BKT transition \cite{Sunami2021}.
	At short and intermediate times the spatial decay of the correlation function is compatible with algebraic scaling, including the effect of inhomogeneity of the system, and with exponential scaling for long times \cite{supp}. 
	This confirms that the transition is indeed of BKT type in time. 
	We identify the crossover time $t_c$, as the time at which the correlation function becomes better described by exponential scaling rather than algebraic; see Fig.\ \ref{fig:corrfunc}B.
	For the dynamic BKT transition, we expect self-similar dynamics with a length scale that depends linearly on time \cite{Comaron2019,Mathey2010,scaling}. 
	Motivated by this, in Fig.\ \ref{fig:corrfunc}B we plot the correlation functions with rescaled length $r'=rt/t_c$ using $t_c \sim 0.5$ s.
	This shows convincingly that the fluctuations in the system only depend on the rescaled parameter $r'$ through a universal function, which we find to be close to the expected power-law function at the equilibrium BKT crossover (Fig.\ \ref{fig:corrfunc}B).
	%At long times, the scale invariance is broken, resulting in deviation from the universal function.
	We find the same behavior independent of the initial condition (temperature) of the system, demonstrating the robustness of the scale-invariant behavior near the critical point \cite{supp}.
	
	At long times scale invariance is broken by vortex excitations, which results in an emergent length scale $\xi \approx 1/\sqrt{n_v}$ where $n_v$ is the vortex density.
	To demonstrate this, in Fig.\ 2C we plot the correlation functions at long times against the rescaled distance $\tilde{r}=r\sqrt{n_v}$ \cite{scaling}.
	We obtain $n_v$ from the occurrence of sudden jumps of phases which indicate the presence of a vortex core \cite{Sunami2021,Hadzibabic2006, supp}.  
	These transformed correlation functions are also time independent, showing that the system is characterized solely by the vortex density above the transition point. 
	
	\begin{figure*}
		\centering
		\includegraphics[width=0.8	\textwidth]{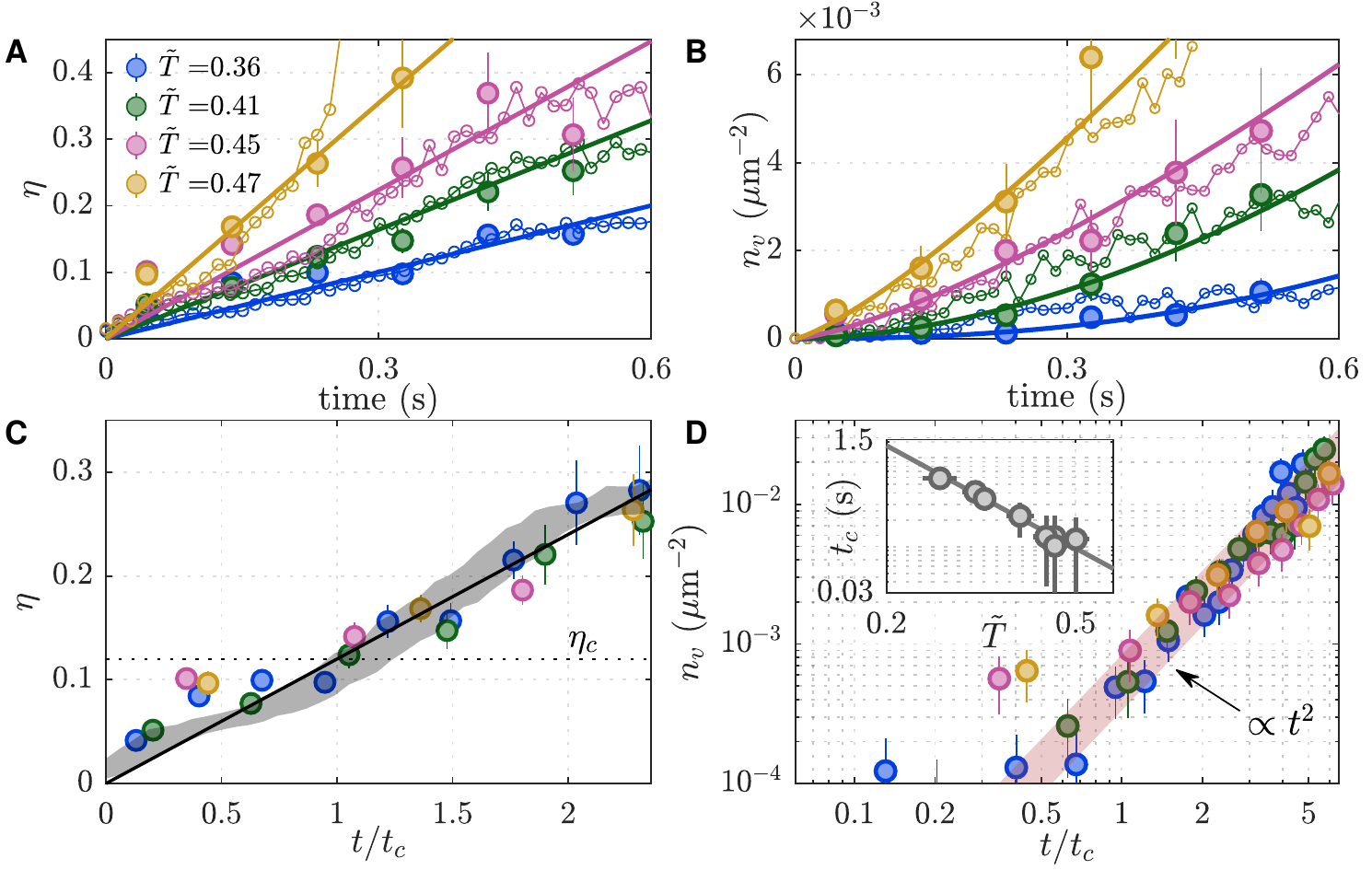}
		\caption{ \textbf{Universal behavior across the dynamic BKT transition.}
			(\textbf{A}) Measured algebraic exponent $\eta (t)$  after the quench at different initial $\tilde{T}$ (filled circles) and the corresponding simulation results (open circles). 
			Solid lines are the linear fit to the experimental data.
			(\textbf{B}) Time evolution of the measured (filled circles) and simulated (open circles) vortex density $n_v(t)$, where the solid lines are the power-law fits to the experimental data.
			(\textbf{C}) Scaled time evolution $\eta(t/t_c)$ according to the $\tilde{T}$-dependent crossover time $t_c$.
			The universal behavior is used to determine the critical exponent $\eta_c =0.13(1)$ (dotted line) at $t/t_c=1$.
			The gray shaded curve corresponds to the simulation result at $\tilde{T}\sim0.4$ and the solid line is a guide to the eye.
			(\textbf{D}) Scaled time evolution of $n_v$, plotted on a log-log scale, displays a universal vortex growth at times above $t_c$, 
			which is in agreement with the RG scaling $n_v \propto t^2$ (shaded region).
			The inset shows the dependence of $t_c$ on $\tilde{T}$ with a solid line as a guide to the eye.
		}
		\label{fig:vortex} 
	\end{figure*}	
	%\newpage
	
	Having verified the behavior of the dynamic BKT transition, we now analyze its universal characteristics by varying $\tilde{T}$. 
	The time evolution of the algebraic exponent $\eta$, determined via an algebraic fit to $C(r)$, exhibits a linear increase where the increase is faster for higher $\tilde{T}$ (Fig.\ 3A). 
	This shows that the dynamics is accelerated at higher $\tilde{T}$ and the system quickly crosses over to the thermal phase. This is also reflected in the measurements of the vortex density $n_v$, showing a faster growth at higher $\tilde{T}$ (Fig.\ 3B). 
	We find the vortex growth to follow a power-law scaling as expected from the RG predictions \cite{supp}.
	We compare the measurements of $\eta$ and $n_v$  with the corresponding results of classical-field simulations which give consistent dynamics (Fig.\ 3,  A and B).

	To confirm universal scaling, we show  $\eta$ and $n_v$ as a function of scaled time $t/t_c$ in Fig.\ 3, C and D. 
	The time evolutions for various initial values of $\tilde{T}$ collapse onto a single curve, showing the robustness of dynamical scaling. 
	We find a linear increase of $\eta$ across $t = t_c$.
	In equilibrium theory, $\eta$ scales approximately linearly with temperature in the superfluid regime, i.e.\ $\eta \propto T/4T_{\mathrm{BKT}}$ \cite{Nelson1977}, thus connecting the temperature scale with phase fluctuations.
	From the linear estimate, we obtain the critical exponent $\eta_c = 0.13(1)$ at $t/t_c=1$, which is below the universal value $\eta_{\mathrm{BKT}}=0.25$, because of the finite-size of the system \cite{Sunami2021}. The linear increase of $\eta$ above $t/t_c=1$ is a precursor of non-equilibrium superheated superfluid \cite{Mathey2017}, which occurs due to a delayed vortex proliferation. 
	%This is in agreement with suppressed vortex density near $t/t_c=1$ in Fig.\ 3D. 
	From the vortex growth above $t/t_c=1$ we obtain universal power-law scaling $n_v \propto t^{2\nu}$ with $\nu \sim 1$, which agrees with the RG prediction; see below.
	
	We now compare the experimental results with predictions based on the real-time RG equations \cite{Mathey2010,Mathey2017}.
	These equations describe the time evolution of parameters characterizing the system from arbitrary non-equilibrium states flowing towards fixed points which represent possible equilibrium states.
	For the dynamic BKT transition, the real-time RG equations are \cite{Mathey2010,Mathey2017,supp}
	\begin{equation}
	\frac{dg}{dt} = \left(2-\frac{1}{2\eta} \right) \frac{g}{t}, \label{eq:flow1}  \\ 
	\end{equation}
	\begin{equation}
	\frac{d\eta}{dt} = \frac{\pi g^2}{16\eta t} + \gamma, \label{eq:flow2}
	\end{equation}
	%The exponent $\eta$ flows due to vortex unbinding and additional, technical heating of the experiment, while the vortex fugacity $g$ decreases for $\eta<1/4$, and increases for $\eta>1/4$. 
	where the vortex fugacity $g$ is related to $n_v$ and healing length $\xi_h = 1/\sqrt{n\tilde{g}}$ \cite{Wen2010,supp}.
	This RG flow derives from the dynamical sine-Gordon model, serving as a dual model for describing the BKT transition and we have added a phenomenological heating
	term $\gamma$ to account for the slow trap-induced heating of the system\cite{supp}.
	For $\eta<1/4$, the fugacity is strongly suppressed, resulting in a linear dispersion $\omega_k = c k$. 
	As $\eta$ increases in time, the vortex fugacity becomes relevant and increases, resulting in a dispersion $\omega_k = c \sqrt{k^2 + 1/\xi^2(t)}$. 
	As argued above, this is indeed supported by the two-step scaling behavior demonstrated above.
	Furthermore, at long times, we have $1/(2\eta) \ll 2$, yielding $n_v \sim g \sim t^2$, as observed in Fig.\ 3D.
	
	\begin{figure}
		\centering
		\includegraphics[width=0.99\linewidth]{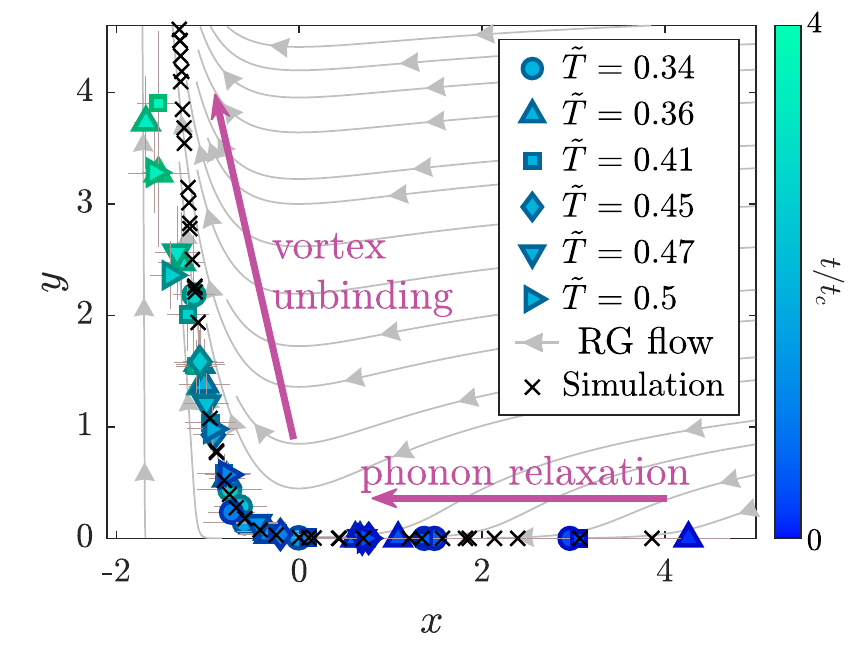}
		\caption{\textbf{Real-time RG flow and measurements.}
			Gray lines with arrows are the RG flow of parameters $x$ and $y$; see text and \cite{supp}.
			These are compared with the experimental data for six different initial temperatures, where the experimental time is scaled by the corresponding $\tilde{T}$-dependent crossover time $t_c$.  
			The results from the numerical simulation at $\tilde{T}\sim0.5$ are shown as black crosses.
		}
		\label{fig:diagram}
	\end{figure}	
	
	In Fig.\ \ref{fig:diagram}, we plot the experimental observations together with the RG flow diagram of Eqs.\ \ref{eq:flow1} and  \ref{eq:flow2}.  
	For this representation, we define $x=1/(2\eta)-1/(2\eta_c)$ and $y=\sqrt{2}\pi g$. 
	This ensures $x_c=0$ at $\eta=\eta_c$ independent of system sizes, where $\eta_c=0.13$ for our finite-sized system and $\eta_c = \eta_{\mathrm{BKT}}=1/4$ is the theoretical predictions in the thermodynamic limit.
	Our results follow a universal trajectory in the flow diagram. 
	The quenched system begins at large $x$, where vortex excitations are suppressed and fugacity is small. 
	Later on, non-equilibrium phonon creation drives the system towards smaller $x$, however still with suppressed $y$. As the system approaches the critical point $x_c = 0$, the onset of vortex excitation drives the transition. 
	
	%In conclusion, we have realized a dynamic BKT transition, induced by splitting a 2D quantum gas in two, thus quenching the system far from equilibrium. 
	%We show that the subsequent dynamics displays a two-step scaling behavior, based on the single-particle correlation function determined by matter-wave interferometry. 
	%The scaling exponent of the vortex density is also extracted from these data. 
	%We demonstrate that the time evolution of these observables is captured by a real-time RG approach, underscoring the universality of the observed dynamics. 
	%On the experimental side, our results demonstrate the versatility of our matter-wave interferometry approach. 
	%Conceptually, our results demonstrate that real-time RG captures the critical dynamics of non-equilibrium many-body systems, suggesting an intriguing perspective on many-body dynamics via the RG.
	
	Our work provides a comprehensive understanding of non-equilibrium dynamics across the BKT transition. The experimental measurements support the real-time RG picture of the universality out of equilibrium indicating that it is an excellent starting point for the theoretical study of a wide range of many-body dynamics within the framework of RG.
	The results also shows that our matter-wave interference technique is ideally suited for further in-depth investigation of universal dynamics in 2D systems, such as the Kibble-Zurek scaling \cite{Kibble1976} and non-thermal fixed points \cite{Schole2012}.
	
	\subsection*{Acknowledgements}
	We acknowledge discussions with Junichi Okamoto on theoretical analysis and thank John Chalker for comments on our manuscript.
	\textbf{Funding:} The experimental work was supported by the EPSRC Grant Reference EP/S013105/1. S. S. acknowledges the Murata scholarship foundation, Ezoe foundation, Daishin foundation and St Hilda’s College, Oxford for financial support. D. G., A. B., A. J. B. and K. L. thank the EPSRC for doctoral studentships. L. M. acknowledges funding by the Deutsche Forschungsgemeinschaft (DFG) in the framework of SFB 925 – project ID 170620586 and the excellence cluster ‘Advanced Imaging of Matter’ - EXC 2056 - project ID 390715994. V.P.S. acknowledges funding by the Cluster of Excellence ‘QuantumFrontiers’ - EXC 2123 - project ID 390837967.
	\textbf{Author contributions:} 
	S.S. performed the experiments and data analysis.
	V.P.S. and L.M. developed numerical and analytical models and contributed to the interpretation of our experimental data. 
	S.S. and V.P.S. wrote the manuscript. 
	L.M. and C.J.F. supervised the project. 
	All authors contributed to the discussion and interpretation
	of our results.
	\textbf{Competing interests:} 
	The authors declare no competing interests.
	%\textbf{Data and materials availability:}
	%All data presented in this paper will be deposited at Zenodo.

	%\bibliography{refs} 
	
	%
	
	\newpage

	\setcounter{equation}{0}
	\setcounter{figure}{0}
	\setcounter{table}{0}
	\renewcommand{\theequation}{S\arabic{equation}}
	\renewcommand{\thefigure}{S\arabic{figure}}
	
	\newpage
	\section*{Supplementary Materials}
	%	Supplementary Text
	%	\\ 
	%	Figs. S1 to S9 \\
	%	References \cite{Luksch2019,Holzmann2010,Bentine2020,Prokofev2002,Hung2011,Hadzibabic2008,Fletcher2015,Holzmann2008,Kruger2007,Boettcher2016,Posazhennikova2006,Kogut1979,Giamarchi2003,Kosterlitz1974,Singh2017}
	%	
	%	\newpage
	%	
	\subsection*{Preparation of non-equilibrium 2D systems}
	
	\baselineskip14.4pt
	
	We begin with an ultracold Bose gas of approximately $9 \times 10^4$ \Rb{87} atoms in the $F=1$ hyperfine ground state, at temperatures $T = $ 40 -- 70 nK loaded adiabatically into a cylindrically symmetric, single-well quasi-2D potential as described in detail in Refs.\ \cite{Harte2018,Bentine2020,Luksch2019}.
	%The quasi-2D condition $\mu, k_BT \lesssim \hbar \omega_z$ ensures the kinetic or interaction energy cannot populate excited states significantly, thereby allowing the factorisation of the `frozen' degree of freedom $z$ from the other degree of freedom $r$.
	The trap is created by a multiple-radiofrequency-dressed potential \cite{Barker2020,Sunami2021} with three RF components ($f_1,f_2,f_3$) = (7.14, 7.2, 7.26) MHz. %80*1.18.
	The static quadrupole field $\bm{B}(\bm{r})=b(x\bm{e}_x+y\bm{e}_y-2z\bm{e}_z)$ has field gradient of $b=\SI{145}{G\, cm}^{-1}$.
	The single-well potential has anisotropic confinement for radial and axial directions $\omega_r/2\pi = 13$ Hz and $\omega_z/2\pi = 1$ kHz, experimentally determined by the measurement of dipole oscillation in the trap.
	This gives the dimensionless 2D interaction strength $\tilde{g} = \sqrt{8\pi} a_s/\ell_0=0.076$, where $a_s$ is the 3D scattering length and $\ell_0=\sqrt{\hbar/(m\omega_z)}$ is the harmonic oscillator length along $z$ for an atom of mass $m$.
	These parameters satisfy a quasi-2D condition $k_B T \lesssim \hbar \omega_z$  for the parameters used in this paper:
	the presence of small excited state populations in the $z$ direction at $\hbar \omega_z \sim k_B T$ results in a small reduction of the 2D interaction strength by $\sim 15\%$ \cite{Holzmann2008} however the BKT critical temperature changes by less than 4 $\%$ as a result \cite{Holzmann2010,Fletcher2015}.
	We perform thermometry of the system before the quench (in equilibrium) in the single-well, by measuring the radial expansion of the far wings of density distribution following the release from the trap \cite{Sunami2021}.
	We use temperature scale $T_0=\sqrt{6N}(\hbar \omega_r/\pi k_B) \approx \SI{140}{\nano \kelvin}$ and report $\tilde{T} = T/T_0$ in the main text.
	
	\begin{figure*}[ht]
		\centering
		\includegraphics[width=0.7\textwidth]{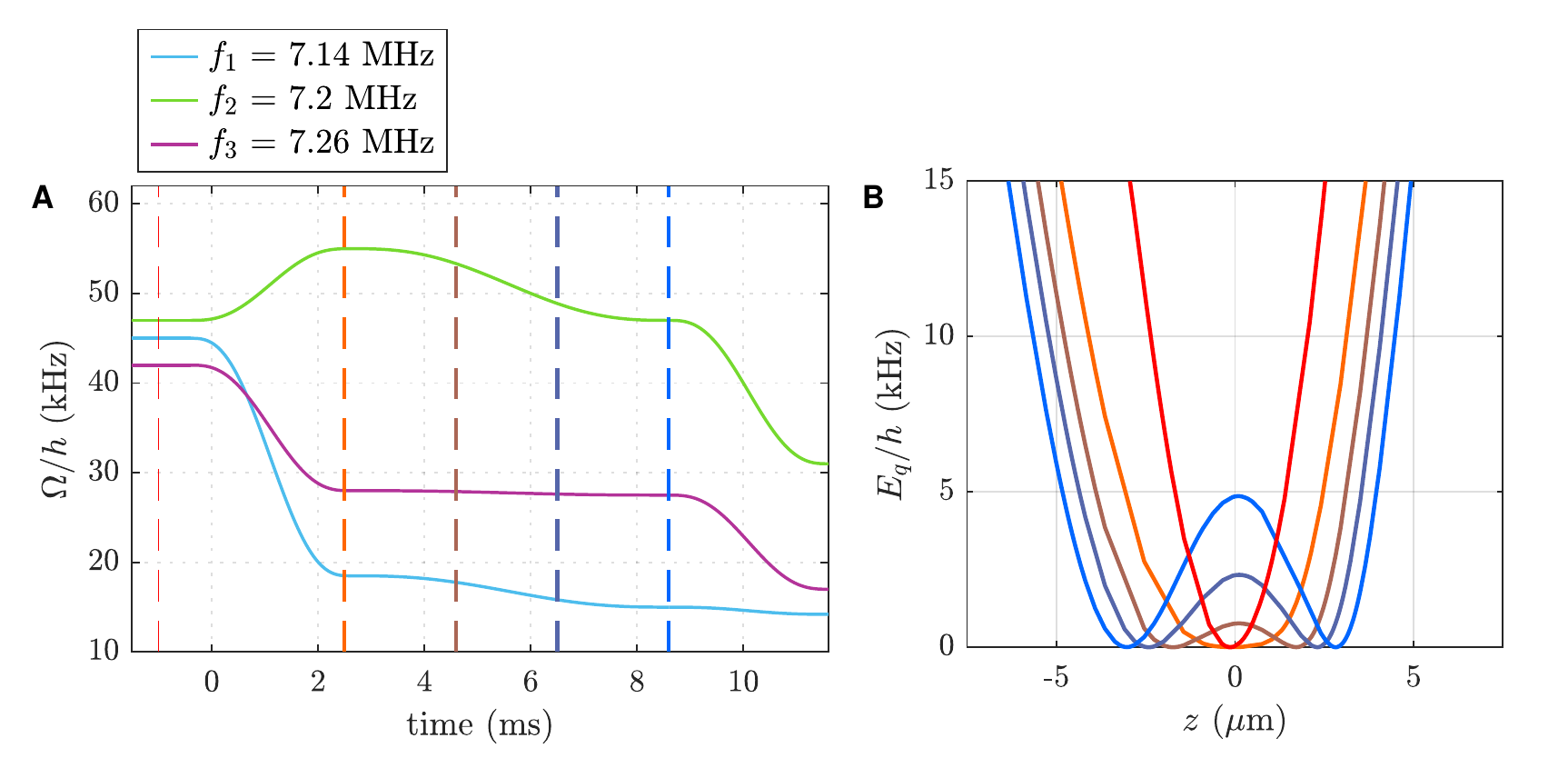}
		\caption{\label{fig:splitting} \textbf{The splitting procedure.} 
			(\textbf{A}) The experimental procedure to transform the potential from single- to double-well. RF amplitudes are expressed in terms of Rabi frequencies $\Omega_i = g_F \mu_B B_i /2\hbar$ for components $i=1,2,3$, where $B_i$ is the RF magnetic field amplitude, $g_F$ is the Land\'e g-factor and $\mu_B$ the Bohr magneton.
			(\textbf{B}) The eigenenergy $E_q$ for atoms in trapped dressed eigenstate $\tilde{m}=1$, which form the trapping potential \cite{Harte2018}.
			We obtain position-dependent eigenenergies $E_q$ using the Floquet numerical simulation \cite{Harte2018,Bentine2020} for the static field gradient $b$ and RF amplitudes $\Omega_i$ for each RF components $i=1,2,3$ and the plotted potentials include the effect of gravity which acts in the direction of $-\bm{e}_z$.
			The red curve indicates the trap before the splitting sequence and the times for other colors are indicated by vertical dashed lines with corresponding colors in the left panel.
			The position $z$ is defined relative to the center of the trap.
			%In the double-well, due to gravitational potential, the two trap minima have small difference in the confinements.
			To realise 2D confinements for the two wells $\omega_z/2\pi=\SI{1}{\kilo\hertz}$, we further modify the trap shape following the complete decoupling of the two wells (9-11.5 ms in left panel).
		}
	\end{figure*}	
	
	After holding the gas for 400 ms in the single trap \cite{Sunami2021,Hadzibabic2006}, we split the cloud into two daughter clouds by changing the vertical trap geometry from single- to double-well potential over 12 ms, thereby splitting the gas as illustrated in Fig.\ 1A and Fig.\ \ref{fig:splitting}.
	This duration is chosen to be much shorter than the typical timescale for the radial dynamics $\sim 2\pi/\omega_r \sim 80$ ms.
	Following the splitting, each well confines $N\sim4.5\times 10^4$ atoms and final vertical trap frequency is $\omega_z/2\pi  = \SI{1}{kHz}$ for both trap minima, satisfying the quasi-2D condition. 
	We ensure equal population of the two wells by maximizing the measured contrast of matter-wave interference patterns.
	Shortly before the splitting, we change the radial trapping frequency to $\omega_r/2\pi = \SI{11}{\Hz}$ by the change of $b$ over 10 ms; 
	this prevents the quench from exciting the monopole mode in the radial direction by matching the radial density profiles before and after the splitting as much as possible.
	This process, as well as the splitting procedure, are performed with duration much longer than characteristic timescale for the atomic motion in the vertical direction, $ \tau_{\mathrm{trap}} \sim 2\pi/\omega_z \sim $ 1 ms such that the system remain 2D.
	Following the splitting, the static quadrupole field has field gradient of $b=\SI{94}{G\, cm}^{-1}$.
	%The characteristic dimensionless 2D interaction strength is $\tilde{g} = \sqrt{8\pi} a_s/\ell_0=0.076$, where $a_s$ is the 3D s-wave scattering length and $\ell_0=\sqrt{\hbar/(m\omega_z)}$ is the harmonic oscillator length along $z$ for an atom of mass $m$.
	The spatial separation of the double-well is $d = \SI{7}{\micro \metre} \gg \ell_0$, which ensures the decoupling of the two clouds for the temperature range and trap parameters used in this work.
	The speed of sound in the 2D Bose gas is given by $c = \hbar\sqrt{n\tilde{g}}/m \sim \SI{1}{\micro \meter \, \milli\second}^{-1}$.
	
	\subsection*{Quench across the critical point}
	With the same trap parameters, interaction strength and similar atom number as the ones used in this work, the equilibrium BKT critical temperature is at $\tilde{T}_{c,\mathrm{eq}} = 0.53(1)$, obtained via interferometric measurement of the first-order correlation functions and vortex densities \cite{Sunami2021}.
	Thus, the initial conditions of the system $\tilde{T}$ is chosen to satisfy $\tilde{T} \lesssim \tilde{T}_{c,\mathrm{eq}}$ and $\tilde{T}_f \gtrsim \tilde{T}_{c,\mathrm{eq}}$, where the value after the quench, $\tilde{T}_f \sim 1.67 \tilde{T}$, is higher due to the change in atom number and radial trapping frequencies.
	$\tilde{T}$ has one-to-one mapping to the peak phase-space density $\mathcal{D} = n\lambda^2$, where $\lambda=h/\sqrt{2\pi m k_B T}$ is the thermal de Broglie wavelength, of the trapped gas as shown below, and thus can be used to identify whether the system should lie in the superfluid regime of the BKT transition if the system is in equilibrium, with $\tilde{T}_{c,\mathrm{eq}} $ being the critical value.
	
	To demonstrate the mapping, we have obtained the theoretical prediction of density distribution in a harmonic trap by the application of classical-field simulation results in Ref.\ \cite{Prokofev2002} for uniform systems to inhomogeneous systems within the local density approximation (LDA).
	The applicability of LDA in this method was confirmed by experiments in Ref.\ \cite{Hung2011} for the range of interaction strengths which includes the value we use.
	To complement the simulation in \cite{Prokofev2002} which was performed only in the fluctuation region around the superfluid critical point, we have used the Hartree-Fock prediction \cite{Hadzibabic2008} deep in the normal regime \cite{Hung2011}.
	These predictions smoothly connect at local phase-space density $\mathcal{D} \sim 2.5$ and give the density distribution.
	
\begin{figure}[h]
		\centering
		\includegraphics[width=0.9	\linewidth]{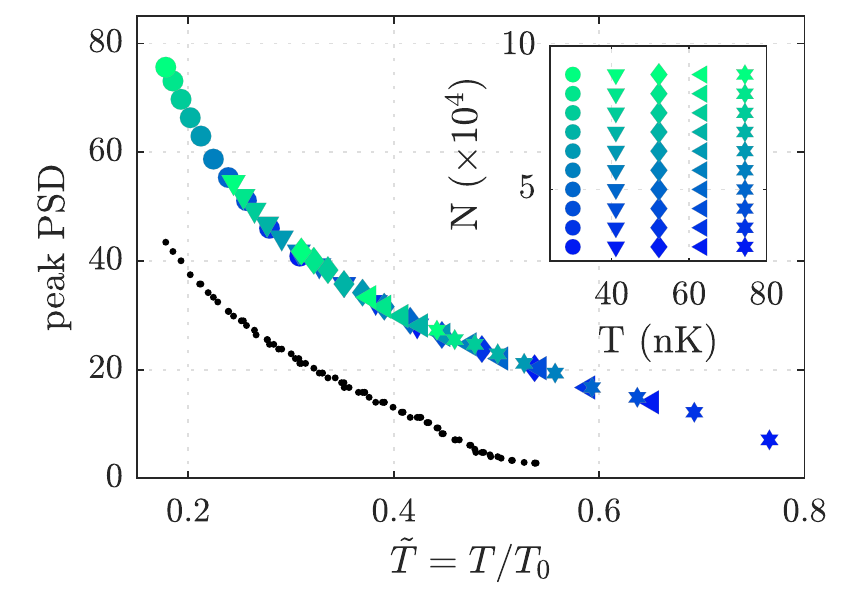}
		\caption{\label{fig:PSD} \textbf{Phase-space density and the reduced temperature $\tilde{T}$.}
			Peak PSD of harmonically-trapped 2D gases in equilibrium for a range of temperature and atom number indicated in inset are plotted for $\omega_r/2\pi = 13$ Hz. 
			The results with $\omega_r/2\pi = 11$ Hz agrees with this curve.
			Black points represent the peak PSD we expect after the quench we described above, from the initial reduced temperature $\tilde{T}$.
		}
	\end{figure}

	In Fig.\ \ref{fig:PSD}, we show the peak PSD for various atom numbers and temperature in 2D harmonic trap. 
	The peak PSD is only dependent on the reduced temperature $\tilde{T}$, supporting the description above.
	At $\tilde{T} = \tilde{T}_{c,\mathrm{eq}} $, the peak PSD is $\sim 20$, as observed in \cite{Sunami2021}.
	The phase diagram in equilibrium (Fig.\ 1B) was also obtained using this method and gives the phase boundary $\tilde{T} = \tilde{T}_{c,\mathrm{eq}} $ on $T-N$ plane.
	%This coincides with the contour for the superfluid fraction of $50 \%$ at which the superfluid covers the entire region of Thomas-Fermi peak that we analyze (see below).
	We note that, as studied in Refs. \cite{Fletcher2015,Holzmann2008}, the ideal-gas condensation, which accompanies the divergence of peak PSD \cite{Holzmann2008}, is suppressed at the interaction strength used in this work and we can neglect its effect on the dynamics \cite{Fletcher2015}. 
	%
	%For the range of parameters described above, the peak phase-space density $\mathcal{D} = n\lambda^2$, where $\lambda=h/\sqrt{2\pi m k_B T}$ is the thermal de Broglie wavelength, changes from \textcolor{red}{$\mathcal{D} \gtrsim 25 $ to $\mathcal{D} \lesssim 20 $} by the quench, bringing the system across the BKT critical point $\mathcal{D}_c \sim 20$ for our trapped system \cite{Sunami2021}.
	
	%\subsubsection*{Equilibrium phase diagram}
	%The phase diagram of 2D Bose gases in the harmonic trap (Figure 1B) was obtained by the application of classical-field simulation results in Ref.\ \cite{Prokofev2002} for uniform systems to the harmonically-trapped system with $\omega_r/2\pi=11$ Hz within LDA, as experimentally demonstrated in Ref.\ \cite{Hung2011} for the range of interaction strengths which includes the value we use, $\tilde{g}=0.076$.
	%To complement the simulation in \cite{Prokofev2002} which was performed only in the fluctuation region around the superfluid critical point, we have used the Hartree-Fock prediction \cite{Hadzibabic2008} deep in the normal regime. 
	%These predictions smoothly connect at local phase-space density $\mathcal{D} \sim 2.5$ and gives the prediction of the total and superfluid density distribution.
	%To obtain the phase diagram shown in Fig.\ 1, we determined the phase boundary for each temperature by the total atom number where the superfluid fraction exceeds 50 $\%$, at which the size of superfluid region covers the central peak of density distribution where we perform the correlation analysis \cite{Sunami2021}.
	%The superfluid transition in harmonically-trapped 2D Bose gas is usually defined as the temperature where the superfluid appears at the center of the cloud (with zero superfluid fraction and hence expected to show no quasi-long-range order). 
	%The prediction of phase boundary obtained from our method agrees with previous theoretical work \cite{Holzmann2010} for this definition at our interaction strength $\tilde{g}=0.076$.
	
	\subsection*{Image analysis}
	After variable time $t$ following the quench, we abruptly turn off the trap and image matter-wave interference patterns with a spatially-selective repumping technique to obtain the local fluctuation of relative phases, as described in detail in Ref.\ \cite{Sunami2021}.
	In this method, we apply a spatially-modulated laser beam that optically pumps a thin slice through the cloud of atoms from $F=1$ hyperfine state to $F=2$, which we subsequently detect by absorption imaging.
	The repumping beam is a thin sheet of thickness $L_y = \SI{5}{\micro \metre}$, which goes through the centre of the cloud and the sheet is normal to the imaging light, as illustrated in Fig.\ 1A.
	
	\begin{figure*}[ht]
		\centering
		\includegraphics[width=0.65 \textwidth]{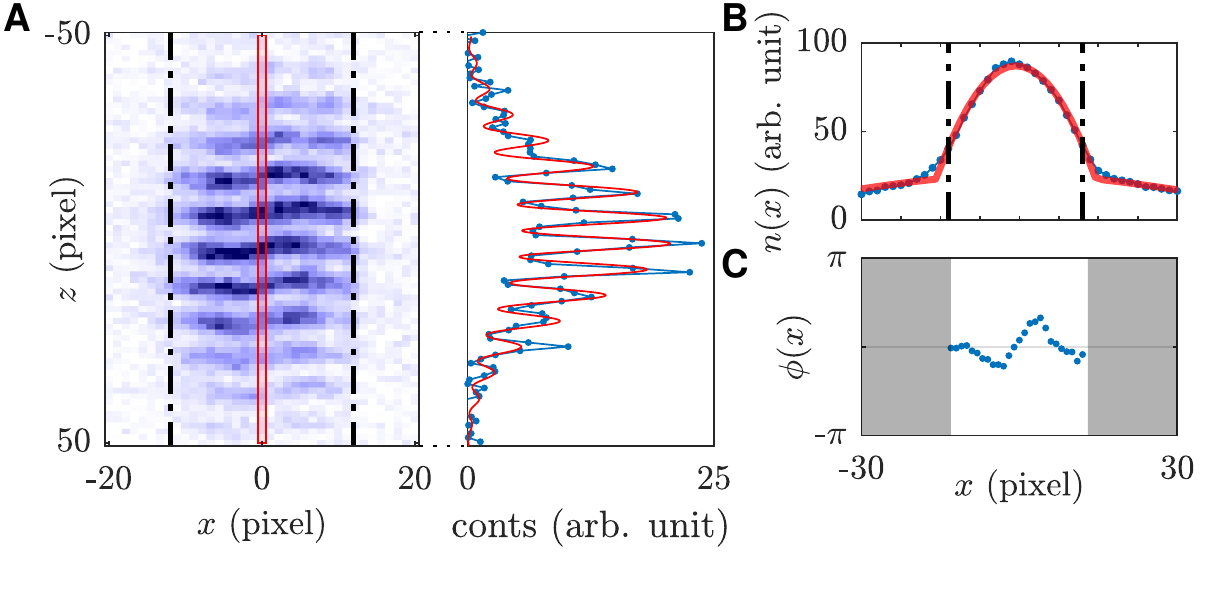}
		\caption{\label{fig:densityFringe} \textbf{Interference analysis and region of interest.} 
			(\textbf{A}) An example of interference image recorded after the quench, where pixel size in the image plane corresponds to $\ell_\mathrm{pixel} = \SI{1.67}{\micro\meter}$.
			The density distribution along the pixel column $x=0$ is shown on the right with the fitted curve (red).
			(\textbf{B}) Density distribution recorded after the TOF, integrated along $z$. 
			Red line denotes the fit with the bimodal model Eq.\ \ref{eq:bimodal}. 
			Dash-dotted lines indicate the 80$\%$ of the central Thomas-Fermi peak, which is also shown in \textbf{A} and corresponds to the region where clear interference fringes are observed.
			(\textbf{C}) Obtained phase profile $\phi(x)$ from the fits at each pixel column $x$. 
			We only analyze the phases within 80$\%$ of the central peak.
		}
	\end{figure*}	
	
	In the images, the density distributions along $x$ (the radial density distribution, obtained by integrating the image along $z$) are bimodal, with a narrow central Thomas-Fermi density profile (inverted parabola) and a broad thermal cloud with Gaussian profile, as observed in equilibrium \cite{Sunami2021,Kruger2007}. 
	It is within the central Thomas-Fermi distribution that interference fringes are clearly observed.
	The observed density distribution, integrated along $z$, is fitted well with the bimodal model 
	\begin{equation}\label{eq:bimodal}
	f(x) = n_{\mathrm{TF}}\max(0,1-r^2/R^2)+n_{\mathrm{Th}}e^{-r^2/2\sigma_n^2}, 
	\end{equation}
	where $n_{\mathrm{TF}},R,n_{\mathrm{Th}},\sigma_n$ are the fit parameters, as shown in Fig.\ \ref{fig:densityFringe}B.
	During the TOF with duration $t_{\mathrm{TOF}} \ll 2\pi/\omega_r \sim$ 100 ms that we use in the experiments, $\sigma_n$ increase due to the ballistic expansion of the thermal component while the Thomas-Fermi peak shows negligible expansion and $R$ stays constant.
	For the density distribution along $z$ direction, we evaluate the phase profile $\phi(x)$ of interference patterns by fitting the column density distributions at each pixel column $x$ with
	\begin{equation}\label{eq:mwi}
	f(z)=n_p \exp(-z^2/2\sigma^2)\left[ 1+ c_0 \cos (kz+\phi) \right],
	\end{equation}
	where $n_p,\sigma,c_0,k,\phi$ are fit variables (see Fig.\ \ref{fig:densityFringe}A; red box in the left panel shows the distribution that is being fitted on the right panel) and we perform the fit at each $x$ within the 80$\%$ of the Thomas-Fermi region of the cloud as illustrated in Fig.\ \ref{fig:densityFringe}A: 
	for each image, we repeat the fitting at varying $x$ (red box shown in Fig.\ \ref{fig:densityFringe}A).
	The obtained phase profiles $\phi(x)$, such as shown in Fig.\ \ref{fig:densityFringe}C, encodes the \textit{in situ} relative phases of two gases along a line that goes through the centre \cite{Sunami2021} and reveals the phase fluctuation in non-equilibrium 2D systems.
	Further details of the phase correlation analysis is described in detail in \cite{Sunami2021}.
	To obtain the atom number $N$, we repeat measurements with a large repumping beam that covers the entire density distribution following TOF.
	The detectivity of the absorption imaging for the measurement of $N$ was calibrated using the known Bose-Einstein condensation critical point of 3D gases, as described in detail in \cite{Sunami2021}.
	
	\subsection*{Phase correlation function}
\begin{figure*}[ht]
		\centering
		\includegraphics[width=0.7 \textwidth]{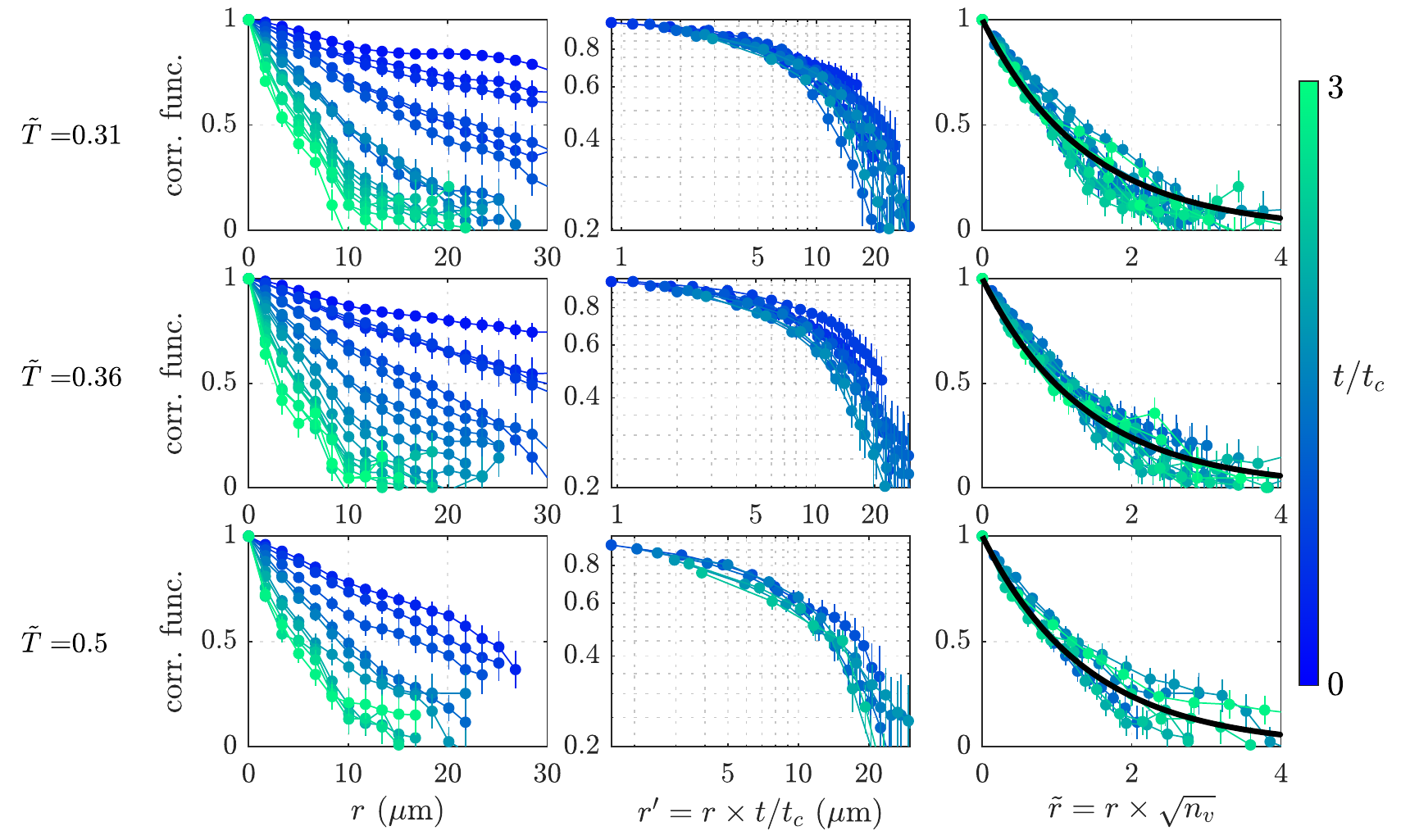}
		\caption{\label{fig:corrFunc} \textbf{Correlation functions and scale-invariant behavior.} 
			The left column shows the measured correlation functions at times ranging from 0 s to $3t_c$, where $t_c = 0.61, 0.52, 0.38$ s are the temperature-dependent crossover time for $\tilde{T}=0.31,0.36$ and $0.5$, respectively.
			Due to smaller region used for the correlation analysis (performed within Thomas-Fermi region which has temperature-dependent size), bottom left panel shows smaller number of data points.
			Middle column shows the rescaled correlation functions in which we perform the rescaling of the distance by $r'=rt/t_c$, in the same way as the main text (Fig.\ 2B).
			We exclude the data at $t< 200$ ms for this plot, as well as in Fig.\ 2B.
			Finally, right panels show the correlation functions in the crossover to thermal regime ($t > t_c$), with distance scaled by the mean vortex-vortex distance $1/\sqrt{n_v}$.
			Black solid line is the exponential function (same as Fig.\ 2C) for guide to the eye.
		}
	\end{figure*}

	At each time and initial condition, we make $N_r=45$ observations of fringes to measure the spatial phase fluctuations. 
	We compute phase correlation function $C(r)$ as defined in the main text.
	In quenched two-dimensional systems, the relationship $C(r)\simeq g_1^2(r)/n^2$ is valid following the so-called light-cone time $t_{lc}= L/2c \sim \SI{15}{\milli \second}$ after the quench, where $L \sim \SI{30}{\micro \metre}$ is the system size and $c\sim \SI{1}{\micro \metre \, \milli \second}^{-1}$ is the speed of sound, as was extensively demonstrated in Refs.\  \cite{Mathey2010,Mathey2017} using numerical simulations.
	The region of interest $w$ is defined so that we analyze the phase data only within 80$\%$ of the Thomas-Fermi peak of the density distribution \cite{Sunami2021}.
	In Fig.\ \ref{fig:corrFunc}, we show the time evolution of correlation functions at three temperatures to demonstrate the robustness of the universal behaviour reported in the main text (Fig.\ 2).
	
	%
	%\newpage
	%\vspace{14cm}
	%\begin{minipage}{1.0\textwidth}
	%\end{minipage}
	%\newpage
	%
	
	The extraction of $\eta$ in the inhomogeneous 2D system relies on the local correlation approximation (LCA) \cite{Boettcher2016}, which is the local density approximation of the correlation properties in the system. 
	We have previously demonstrated the applicability of LCA on the phase correlation function of 2D Bose gases in equilibrium, in Ref.\ \cite{Sunami2021} by comparing to experimentally observed phase correlation functions.
	Essentially, the LCA amounts to the replacement of $\eta$ with a position-dependent one, resulting in the first-order correlation function in the superfluid regime of the form $g_1(\bm{r},\bm{r}') \propto |\bm{r}-\bm{r}'|^{-\eta n_0/\sqrt{n(\bm{r})n(\bm{r}')}}$ where $n(\bm{r})$ is the 2D density and $n_0$ is the peak density.
	For comparison with the averaged phase correlation function, we replace the 2D density with  $n(r) = \langle \sqrt{n(x-r/2)n(x+r/2)} \rangle_{x\in w}$. 
	We then fit the correlation function $C(r)$ with $f_{\mathrm{alg}}(r) = a r^{-2\eta \alpha(r)}$ with $\alpha(r) = \frac{\max(n(r))}{n(r)}$ where $a$ and $\eta$ are fit parameters. 
	The obtained $\eta$ represents the averaged value within the region of interest \cite{Sunami2021}.
	We use this model to obtain the expected correlation function at the crossover in equilibrium, shown in Fig.\ 2B, for $\eta= \eta_c = 0.13$ and assuming Thomas-Fermi density distribution of the form $n(r)=\max(0,1-r^2/R^2)$ with $R=\SI{25}{\micro \metre}$ which is close to the experimental values of $R$.

	We further fit $C(r)$ with $f_{\mathrm{exp}}(r)= b e^{-2r/\xi}$, which models the correlation function decay in the thermal regime of the BKT transition;
	in the thermal regime, the correlation length is typically shorter than the slow variation of inhomogeneous density within the analysis region and we expect almost purely exponential behaviour as we have shown using numerical simulation in Ref.\ \cite{Sunami2021}.
	Since $\xi$ cannot exceed the system size, the value of $\xi$ is bounded by the TF diameter, the approximate system size where BKT physics is observed \cite{Posazhennikova2006}.
	From fits of measured $C(r)$ with $f_{\mathrm{alg}}$ and $f_{\mathrm{exp}}$ and the uncertainties of the data points, we obtain the $\chi^2$ statistic which describes the goodness of fit; lower $\chi^2$ values indicate better fit.
	$\chi^2$ values that are too low, such as $\chi^2 < 4$ for the degree of freedom for the fits performed in this work, indicate incorrect estimation of errors rather than a better fit however the $\chi^2$ values we obtained are comfortably above this threshold.
	At short times, $\chi^2_{\mathrm{alg}}$ is lower than $\chi^2_{\mathrm{exp}}$ however as the system evolves towards the thermal regime, $\chi^2_{\mathrm{exp}}$ becomes lower than $\chi^2_{\mathrm{alg}}$, as shown in Fig. 2B inset.
	We determine the crossover time, at which the correlation function become better described by the exponential model, from the crossover of $\chi^2_{\mathrm{alg}}$ and $\chi^2_{\mathrm{exp}}$.
	For the typical degree of freedom for the fit procedure, a model is accepted if $\chi^2 \lesssim 30$ at 5$\%$ level of significance.
	The crossover typically occur in the range $\chi^2 \sim 20-30$ and algebraic model is accepted for times $t \lesssim t_c$ while exponential model is accepted at long times $t \gtrsim t_c$, with narrow crossover regime around $t\sim t_c$ where both models are accepted.
	
	%\textcolor{red}{The predicted correlation function in equilibrium crossover, in Fig.\ 2B, is obtained as follows ...}%. on function in equilibrium crossover, in Fig.\ 2B, is obtained as follows. 
	%Firstly, we invoke LCA and predict the power-law correlation function in the presence of inhomogeneous density with parabolic profile \cite{Boettcher2016}, $C(r)\propto r^{-\eta_c \max(n(r))/n(r)}$ with $n(r)=\max(0,1-r^2/R^2)$ where we used experimentally determied values of $\eta_c = 0.13$ and TF radius $R=\SI{20}{\micro \metre}$.
	%We further incorporate the effect of finite imaging resolution which results in a small shift of correlation function at short length scale \cite{Singh2014}.

	\subsection*{Consistency of scaling in Fig.\ 2B and Fig.\ 3C}
\begin{figure}[h]
		\centering
		\includegraphics[width=0.8	\linewidth]{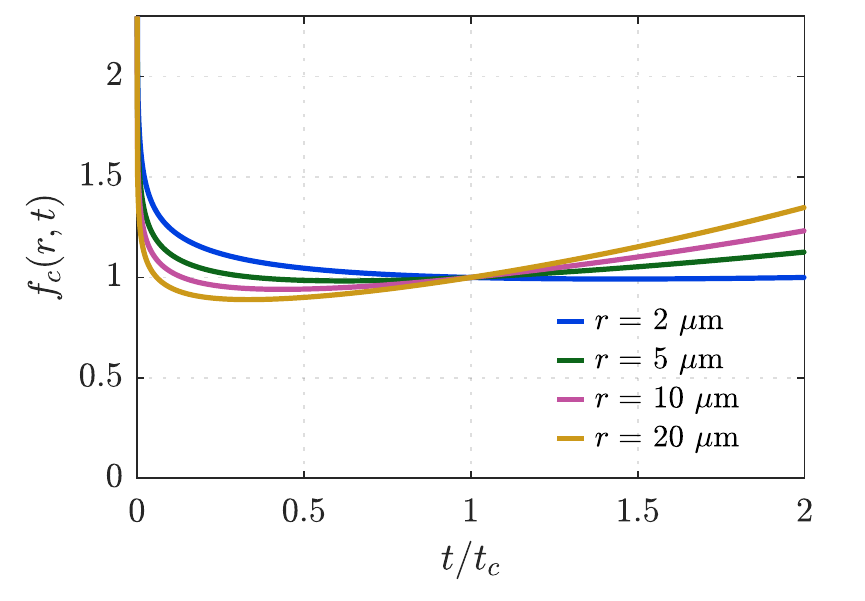}
		\caption{\label{fig:alpha} \textbf{Consistency of the scaling in the power-law regime.}
			The time evolution of the ratio $f_c(r,t)$ for times $t/t_c \in [0,2]$ at various lengths $r$.
		}
	\end{figure}

	According to Fig.2B, within the scale-invariant regime the correlation function $C(r,t)$ should have the form
	\begin{equation}
	C(r,t) \sim C_1(r') = (r')^{-\eta_c},
	\end{equation}
	up to $t \sim 2 t_c$, where $r'=rt/t_c$. At the same time, the linear increase of $\eta$ in Fig.\ 3C implies that 
	\begin{equation}
	C(r,t) \sim C_2(r,t) = r^{-\eta_c t/ t_c}.
	\end{equation}
	To show that these two expressions are consistent, we show in Fig.\ \ref{fig:alpha} the time evolution of 
	\begin{equation}
	f_c(r,t) = \frac{C_1(r')}{C_2(r,t)} = r^{\eta_c (t/t_c - 1)} \Big(\frac{t}{t_c}\Big)^{-\eta_c},
	\end{equation}
	for $t/t_c \ \in [0,2]$. 
	This shows that $f_c(r,t)$ is close to unity for the range of $r$ relevant for our experiment, thus confirming the consistency of scaling demonstrated in Figs.\ 2B and 3C.

	%\subsubsection*{\textcolor{red}{Relaxation of relative and common phase modes following a quench} }
	%The extraction of the first-order correlation function through the interference patterns (relative phases) assumes independent phase fluctuations in the two gases.
	%While this holds for decoupled systems in equilibrium, as studied in Ref.\ \cite{Sunami2021}, the quench scheme result in finite correlation of the phases of two clouds at the time of the splitting since the two clouds originate from the same cloud at $t=0$. 
	%Nevertheless, in 2D bose gases the relaxation of the relative and common phase degrees of freedom occurs rapidly, in contrast to the integrable 1D systems where these two degrees of freedom stay out of equilibrium for extended period of time \cite{Langen2016}. 
	%Specifically, the relative and common degrees relax to achieve complete decoupling of the two wells within the so-called light-cone time $t_{lc}= L/2c \sim \SI{15}{\milli \second}$ where $L \sim \SI{30}{\micro \metre}$ is the system size and $c\sim \SI{1}{\micro \metre/\milli \second}$ is the speed of sound, as was extensively demonstrated in Refs.\  \cite{Mathey2009,Mathey2010,Mathey2017} using numerical simulations.
	%This timescale is significantly shorter compared to the dynamics we have reported in the main text, and we thus safely neglect the effect of the initial condition on the relative phase fluctuations.
	%\textcolor{red}{show absolute phase histogram to show the decoupling of the wells?}
	
	\begin{figure*}[ht]
		\centering
		\includegraphics[width=0.7	\textwidth]{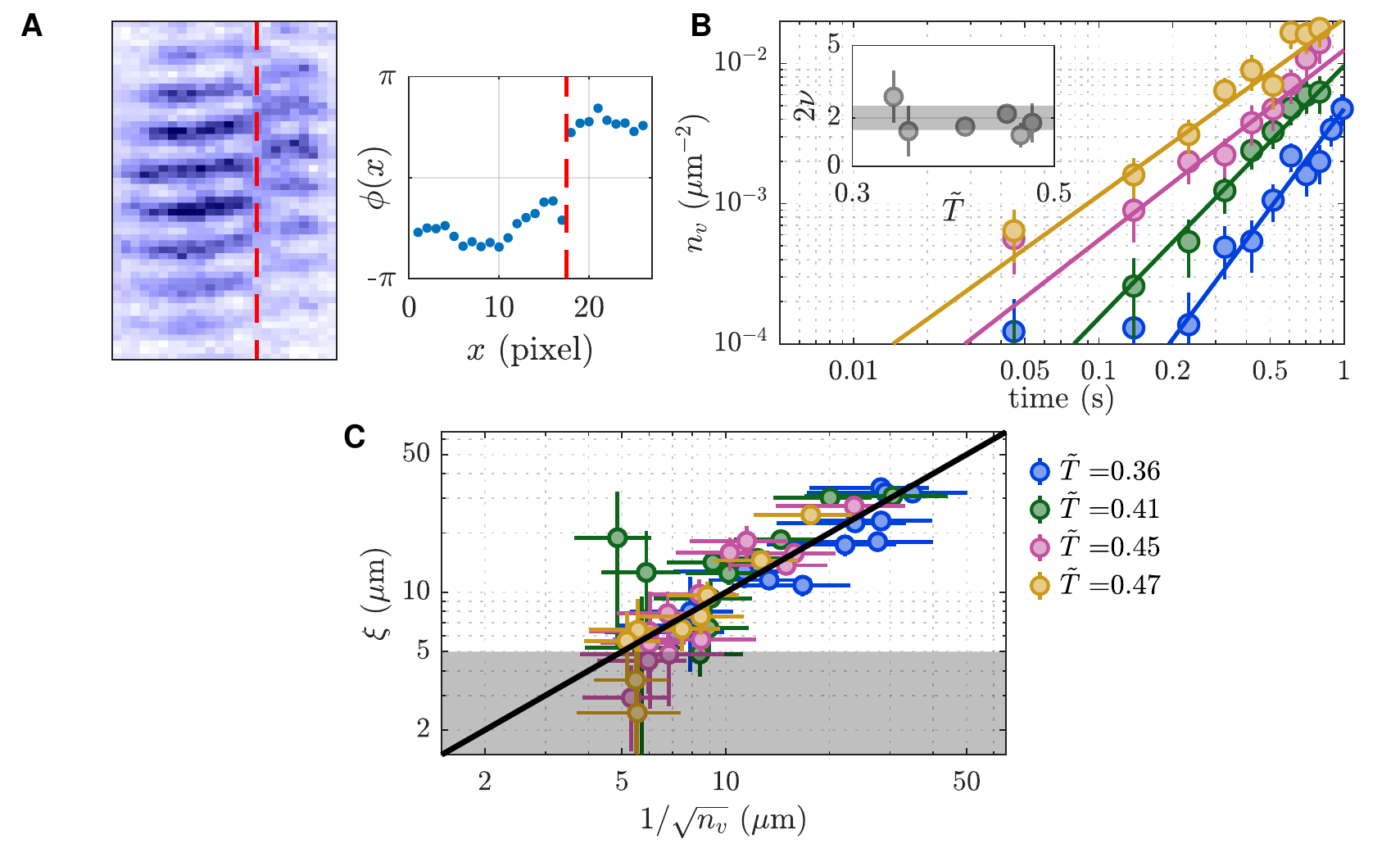}
		%\captionof{figure}
		\caption{\label{fig:corrLength} \textbf{Vortex detection and correlation length.}
			(\textbf{A}) Vortices are detected by the sudden jump of phases. 
			An example interference image with a vortex is shown, where a red dashed line indicates the location of a vortex. 
			Right panel shows the obtained phase distribution $\phi$.
			(\textbf{B}) Time evolution of the vortex density $n_v$ (circles) and the corresponding power law fit $t^{2 \nu}$ (continuous line) are shown for different initial $\tilde{T}$, which are the same as in Fig.\ 3B.
			Inset shows the obtained exponent $2\nu$ and shaded region is the guide to the eye.
			%(\textbf{C}) Time evolution of correlation length $r_0$ obtained by fitting correlation functions with exponential model.
			%Error bars denote standard errors and black shaded region denote  $r_0 < \SI{5}{\micro \metre}$ where the extraction of $r_0$ become unreliable due to the rapid decay of $C(r)$ towards 0 within a few datapoints.
			%(\textbf{D}) Universal dynamics of correlation length.
			%Red shaded region is the guide to the eye with power-law decrease $\propto t^{-1}$.
			(\textbf{C}) Correspondence of the correlation length $\xi$ and the vortex density. 
			The black solid line is the predicted relationship $r_0  = 1/\sqrt{n_v}$ \cite{Wen2010}.
			The black shaded region denotes  $r_0 < \SI{5}{\micro \metre}$ where the extraction of $r_0$ becomes unreliable due to the rapid decay of $C(r)$ towards 0 within a few datapoints.
		}
	\end{figure*}	
	%\end{minipage}
	%\newpage
	
	\subsection*{Vortex detection}
	The method to obtain the vortex density from the interference patterns is described in detail in Ref.\ \cite{Sunami2021}, which is improved upon the method used in Ref.\ \cite{Hadzibabic2006} and takes advantage of the selective imaging method to obtain the vortex density $n_v$ as demonstrated in equilibrium by comparing to the classical-field simulation \cite{Sunami2021}.
	We look for sudden jump of phase $\phi(x)$, and obtain $n_v$ from their occurrences in $N_r=45$ experimental repeats.
	%We first obtain the probability to find a vortex $p_v(t)$ in the phase profile at each time after the quench, by looking for the phase jump above $\pi2/3$.
	%The density slicing method gives the fixed detection region of vortices with area $S=\ell_p L_y$ where $\ell_p= \SI{1.67}{\micro \metre}$ is the image-plane pixel size and $L_y=\SI{5}{\micro \meter}$ is the density slice thickness (see main text, Fig.\ 1A).
	%We thus compute the vortex density by dividing the vortex density by the detection region area: $n_v(t) = p_v(t)/2S$ where the factor of two in the denominator reflects the fact that detected vortices can be in either of two layers.
	In Ref.\ \cite{Sunami2021}, we found good agreement of the vortex density in equilibrium across the BKT transition, with the values obtained from the extensive classical-field predictions performed with the same parameters as in the experiment.
	To confirm this further for the non-equilibrium case, we plot in Fig.\ \ref{fig:corrLength} C the correlation length $r_0$ against $1/\sqrt{n_v}$; 
	since the mean vortex distance $1/\sqrt{n_v}$ determines the correlation length in thermal regime, we expect $\xi = 1/\sqrt{n_v}$.
	The experimental data points are consistent with this prediction, further confirming our vortex detection method. 
	In Fig.\ 3D, we set the lower bound of the vertical axis at $10^{-4} \SI{}{\micro \metre}^{-2}$. This is because observing only a single vortex in the dataset typically result in $n_v \sim  0.9 \times 10^{-4} \SI{}{\micro \metre}^{-2}$ where statistical uncertainty is large and $n_v$ fluctuates between zero and finite value.
	%\textcolor{blue}{new fig: divergence of eta in long time and explosion of error bar.}
	\\
	%\begin{minipage}{1.0\textwidth}

	\subsection*{Real-time RG equations}
	In Ref.\ \cite{Mathey2010,Mathey2017}, the dynamical sine-Gordon model of the form
	\begin{equation}
	\mathcal{L} \sim \frac{\tau}{8 \pi} \Big[ - \frac{(\partial_t \theta)^2}{2 c^2} + \frac{(\partial_x \theta)^2}{2} \Big] + \frac{g}{a^2} \cos\theta
	\end{equation}
	was studied, as a dual model describing the BKT transition. 
	As derived in Ref.\ \cite{Mathey2010}, the dynamical renormalization group (RG) equations are
	%(now the eta + g equations with the flow parameter \ell.)
	\begin{align}
	\frac{dg}{d\ell}&= \left(2-\frac{2}{\tau}\right)g,   \label{eq:mathey1} \\
	\frac{d\tau}{d\ell}&=\frac{\pi^2 g^2}{\tau}. \label{eq:mathey2} 
	\end{align}
	These equations describe the relaxation dynamics of the system at long times. 
	$\ell$ is the flow parameter, related to real time by $t=t_0e^\ell$ and non-universal constant of the RG equation in \cite{Mathey2017} is set so that the numerical prefactor for Eq.\ \ref{eq:mathey2} is unity. 
	These flow equations coincide with the flow equations for the static system in equilibrium, see e.g. \cite{Kogut1979,Giamarchi2003}. 
	While the flow equations of the static system identify which ordered state the system forms, and what the critical properties of the equilibrium phase transition are, the dynamical flow equations describe the universal many-body dynamics across the transition. 
	Written in terms of the time $t$ and $\eta=\tau/4$, they are
	%(eta and g equations with t as the flow parameter)
	\begin{align}
	\frac{dg}{dt}&= \left(2-\frac{1}{2\eta}\right)\frac{g}{t},   \label{eq:matheyt1} \\
	\frac{d\eta}{dt}&=\frac{\pi^2 g^2}{16\eta t}. \label{eq:matheyt2} 
	\end{align}
	To motivate these flow equations, we consider the quantity $\langle |\theta_{k, 0}|^2 \cos^2{\omega_k t} \rangle$, which contributes to the fluctuations of the real-valued field $\theta(r, t)$, the dual field describing the BKT transition in the sine-Gordon model \cite{Mathey2010}. 
	For a linear spectrum $\omega_k = c k$, and for times $t \ll 1/(c k)$, the quantity acts like a static quantity with its initial value.  
	For $t \gg 1/(c k)$, the quantity is dephased to a new near-static quantity, that acts as a slowly varying correction for the low-energy modes with $c k \ll 1/t$. This motivates the $z=1$ scaling that keeps $c k t$ invariant. 
	The inverse time $1/t$ acts as a cut-off on the mode dynamics of the field, which motivates the analogy to equilibrium renormalization flow, in which a momentum cut-off is lowered to improve the low-energy model. 
	At the time $t$, the modes with $\omega_k \approx 1/t$ undergo the dephasing dynamics, acting as a renormalization on the other degrees of freedom. 
	The non-linear term $\sim g \cos \theta$ takes the modes that undergo the dephasing into account. 
	For $\eta$ increasing above the critical value, the term becomes relevant, generating an additional length $\xi$, that enters the dispersion as $\omega_k = c \sqrt{k^2 + 1/\xi^2(t)}$. 
	This dynamical emergence of the length scale $\xi$ indicates the dynamical phase transition.

	\begin{figure*}[ht]
		\includegraphics[width=0.67 \textwidth]{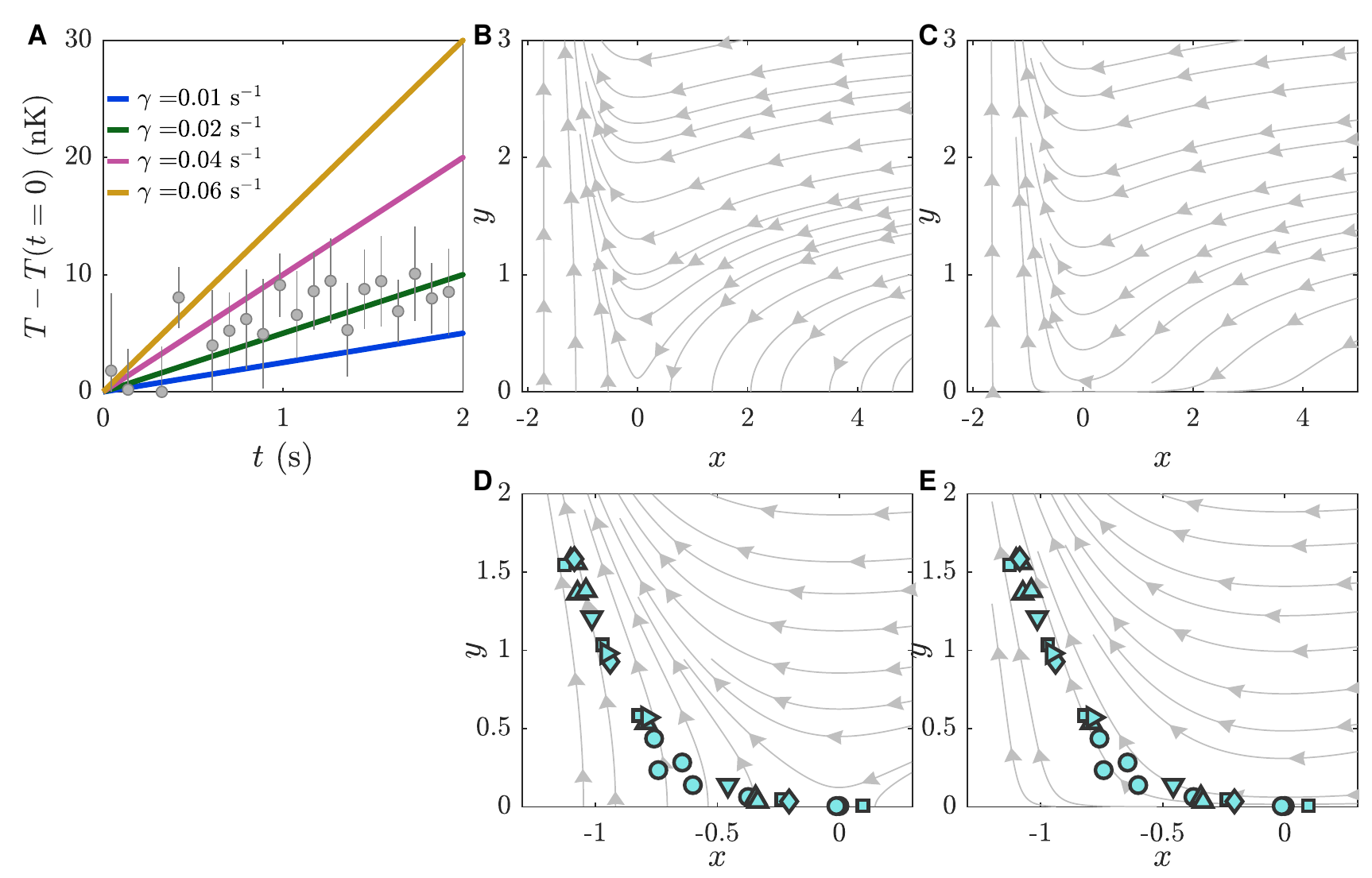}
		\caption{\label{fig:heating} \textbf{The effect of the finite heating in the RG flow.} \
			(\textbf{A}) We find the heating induced by the term $\gamma$ by solving Eq.\ \ref{eq:matheygamma2} with $\eta(t=0)=0$ and assuming $g\ll1$, at different values of $\gamma$. 
			The points represent the measured heating in the trap.
			\textbf{(B,C)} The RG flow diagram plotted for $\gamma=0$ (B) and $\gamma=\SI{0.02}{\second}^{-1}$ (C).
			\textbf{(D,E)} RG flow diagram for $\gamma=0$ and $\gamma=\SI{0.02}{\second}^{-1}$, together with experimental data shown in Fig.\ 4.
		}
	\end{figure*}

	The time evolution of the vortex fugacity is suppressed if the scaling exponent $\eta$ is smaller than the critical value $\eta_{\mathrm{BKT}}$, and increases rapidly if $\eta > \eta_{\mathrm{BKT}}$. 
	As mentioned in the main text, a finite-size system might be characterized by a modified value of $\eta_c < \eta_{\mathrm{BKT}}$.
	The magnitude of $\eta$ is increased by vortex-antivortex unbinding, corresponding to the $g^2$ contribution. 
	As mentioned in the main text, we introduce a phenomenological heating rate $\gamma$, to model the heating due to technical noise:
	%(equations of eta and g, as a function of t, including gamma. This is the set of equations I propose to put in the main text)
	\begin{align}
	\frac{dg}{dt}&= \left(2-\frac{1}{2\eta}\right)\frac{g}{t},   \label{eq:matheygamma1} \\
	\frac{d\eta}{dt}&=\frac{\pi^2 g^2}{16\eta t} + \gamma, \label{eq:matheygamma2} 
	\end{align}
	In the analysis presened in the main text, we use $x=1/(2\eta)-2$ and $y=\sqrt{2}\pi g$ which is the similar form as used by Kosterlitz in Ref.\ \cite{Kosterlitz1974} for equilibrium RG theory of BKT transition.
	The resulting RG equations are
	%(equations in terms of x and y. Please note that the gamma term will now look differently. It shoud have some 1/eta^2 as a prefactor. Please check that the Fig 4 is only weakly affected.)
	\begin{align}
	\frac{dy}{dt} &= -\frac{xy}{t},   \label{eq:dydt} \\
	\frac{dx}{dt} &= % \left(-\frac{1}{8\eta^2}\right) \frac{d\tau}{dt} \nonumber \\
	- \frac{(x+2)^3y^2}{8t} - 2\gamma(x+2)^2. \label{eq:dxdt}
	\end{align}
	For $\gamma=0$, we find the conserved quantity $x^2-y^2=\mathrm{const.}$ near the critical point $x\sim0$, which allowed the visual inspection of critical behavior in Ref.\ \cite{Kosterlitz1974}. 
	The phenomenological term $\gamma$, added in Eq.\ \ref{eq:matheygamma2}, models the finite heating of the system in the trap. 
	In Fig.\ \ref{fig:heating} A, we compare the measured heating in the system with the expected heating in the theoretical model due to our phenomenological term $\gamma$: 
	from Eq.\ \ref{eq:matheygamma2} with the assumption of $g\ll 1$, we find expected temperature of the system using the linear relationship of $\eta$ and system temperature in the superfluid regime $\eta \sim \zeta T$ where $\zeta=4\times 10^{-3}\  \SI{}{\nano \kelvin}^{-1}$, observed in equilibrium with the same trap parameters, interaction strength and similar atom number \cite{Sunami2021}.
	We find reasonable agreement of the measured heating in the trap with the parameter that we use, $\gamma= \SI{0.02}{\second}^{-1}$.
	In Fig.\ 4, to incorporate the finite-size effect which shifts the critical algebraic exponent, we use $x=1/(2\eta) - 1/(2\eta_c)$ with $\eta_c= \eta_{\mathrm{BKT}}=1/4$ for theoretical curves and $\eta_c = 0.13$ for experiments, which results in $x_c=0$ at $\eta=\eta_c$ independent of the system size.
	
	%We use $\gamma = \SI{1}{\second}^{-1}$ which gives the heating close to the experimentally measured value of $\sim \SI{4}{\nano \kelvin}$, as illustrated in Fig.\ \ref{fig:heating} A.

	\begin{figure}[h]
		\centering
		\includegraphics[width=0.8	\linewidth]{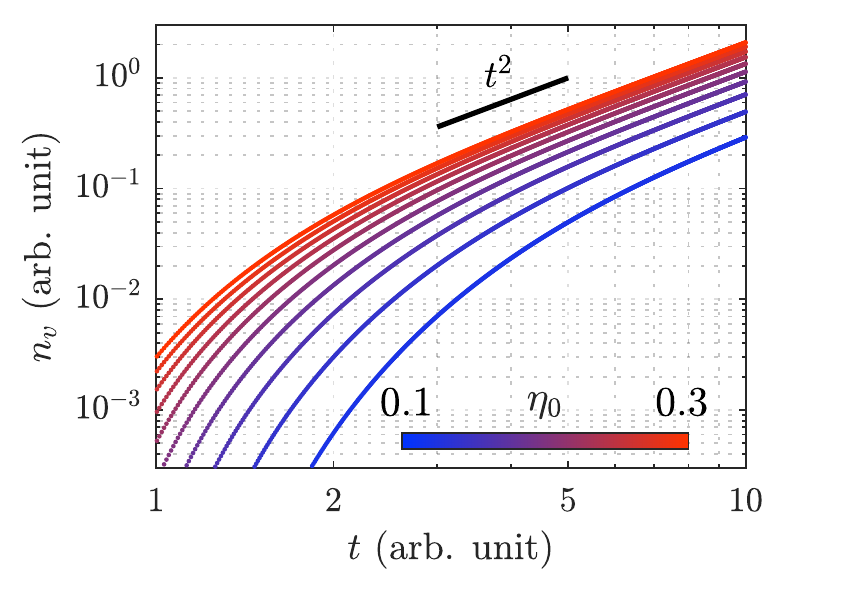}
		\caption{\label{fig:rgtheory} \textbf{Vortex scaling.}
			Time evolution of $n_v$ obtained by numerical integration of the real-time RG equations \ref{eq:matheyt1}, \ref{eq:matheyt2}.
			The slope of solid black line indicates the power-law scaling $t^2$ which all the curves follow at long time.
		}
	\end{figure}	
	
	\begin{figure*}[ht]
		\centering
		\includegraphics[width=0.65	\textwidth]{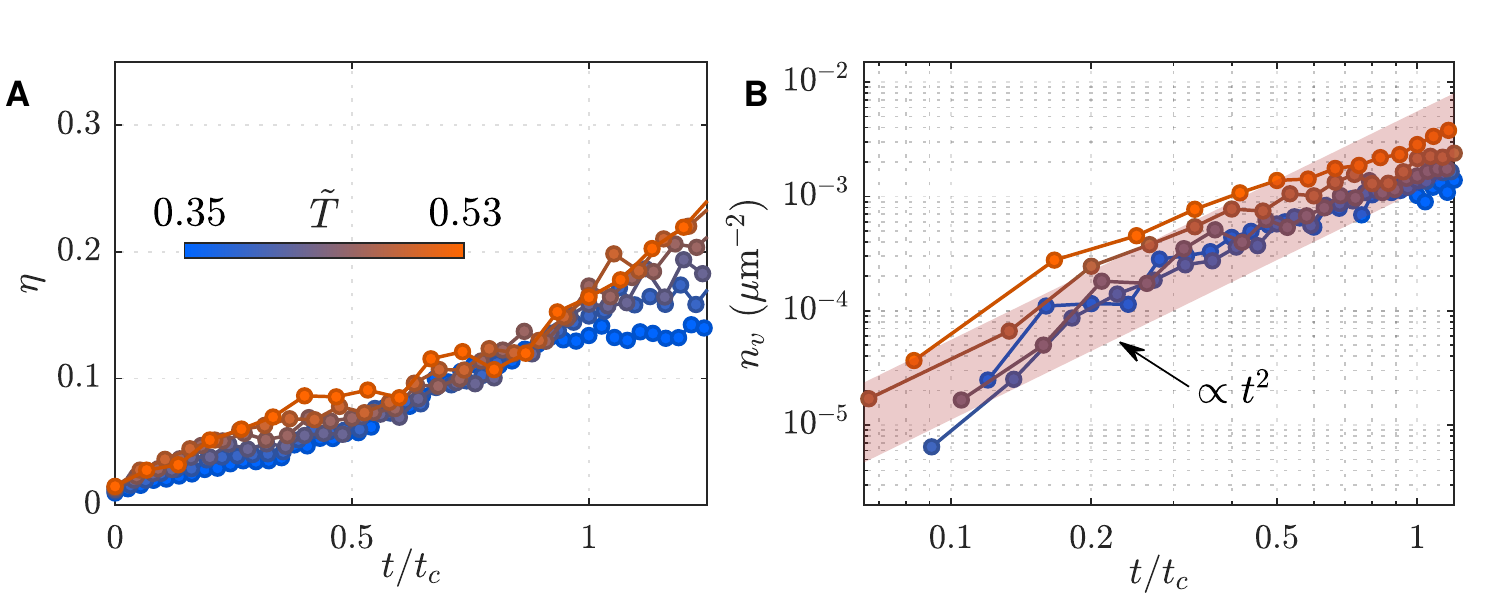}
		\caption{\label{fig:simScaling} \textbf{Universal scaling from numerical simulation.} 
			(\textbf{A}) Time evolution of algebraic exponent against rescaled time $t/t_c$, with temperature-dependent crossover time $t_c$.
			%\textcolor{red}{why only up to t/tc=1??}
			(\textbf{B}) Time evolution of vortex density $n_v$.
			%\textcolor{red}{why deviation, and why not t2 scaling. show tc vs T, or nu vs T? heating included???}
			% We find finite saturation at longer time due to the 
		}
	\end{figure*}	

	The advantage of direct comparison on the RG flow diagram, as performed in Fig.\ 4, is that no concrete initial conditions and timescales are required to compare the theoretical predictions with the experimental findings.
	To obtain Fig.\ 4, we find the vortex fugacity $g$ using \cite{Wen2010}
	\begin{equation}\label{eq:nv}
	n_v(t) = \xi_h^{-2} \exp \left(\frac{2\ln (g/2 )}{2-1/(2\eta)}\right),
	\end{equation} 
	%\textcolor{red}{can we keep 2 here or is it 1/2etac?}
	where $\xi_h \sim \SI{1}{\micro \meter}$ is the healing length of the system which we obtain from the mean density of the system $n$ by $\xi_h=1/\sqrt{n\tilde{g}}$.
	From Eq.\ \ref{eq:nv}, $n_v \propto g$ for large $\eta$ and thus supports $n_v \sim g \sim t^2$ in the main text.
	In Fig.\ \ref{fig:heating} B and C, we compare the RG flow diagram with $\gamma=0$ and $\gamma=\SI{0.02 }{\second}^{-1}$.

	\subsection*{Scaling for vortex unbinding dynamics}
	To demonstrate the predicted scaling of the vortex unbinding dynamics $n_v \propto t^2$, we first numerically solve Eqs.\ \ref{eq:mathey1} and \ref{eq:mathey2} and obtain the results for the time evolution of vortex density.
	We plot the results in Fig.\ \ref{fig:rgtheory} for initial conditions $\eta_0 \in [0.1, 0.3]$ with initial vortex fugacity $g_0 = 0.01$ \cite{Mathey2010}. 
	We find that the increase of $n_v$ follows $t^2$ scaling at long times, independent of initial conditions.

	\subsection*{Classical-field simulations}
	
	We simulate the dynamics of two-dimensional (2D) Bose gases using the classical-field method of Ref. \cite{Singh2017}. 
	The initial system is described by the Hamiltonian 
	\begin{align} \label{eq_hamil}
	\hat{H}_{i} &= \int d \br \, \Big[\frac{\hbar^2}{2m} \nabla \hat{\psi}^\dagger({\bf r})  \cdot \nabla \hat{\psi}({\bf r}) 
	+ V(r) \hat{\psi}^\dagger({\bf r})\hat{\psi}({\bf r})    \nonumber \\
	& \quad  +  \frac{g_{\mathrm{2D}}}{2} \hat{\psi}^\dagger({\bf r})\hat{\psi}^\dagger({\bf r})\hat{\psi}({\bf r})\hat{\psi}({\bf r})\Big],
	\end{align}
	where $\hat{\psi}$ ($\hat{\psi}^\dagger$) is the bosonic annihilation (creation) operator,  $m$ is the atomic mass, and $g_{\mathrm{2D}} = \tilde{g}\hbar^2/m$ is the 2D interaction parameter. 
	%$\tilde{g}= \sqrt{8 \pi} a_s/\ell_z$ is the dimensionless interaction, with $a_s$ being the 3D s-wave scattering length and $\ell_z$ being the harmonic-oscillator length of the confining potential in the transverse direction.
	$V(r)= m\omega_r^2 r^2/2$ describes the harmonic trap potential, where $\omega_r$ is the trap frequency and $r= (x^2 + y^2)^{1/2}$ is the radial coordinate. 
	We choose the total atom number $N= 9 \times 10^4$, $\omega_r/(2\pi) = 13\,  \mHz$ and $\tilde{g}= 0.076$, which are the same as the experiments. 
	For numerical simulations we discretize space on a lattice with discretization length $l=0.5\, \mum$.
	In our methodology we replace the operators $\hat{\psi}$ in Eq. \ref{eq_hamil} and in the equations of motion by complex numbers $\psi$.
	We sample the initial states from a grand-canonical ensemble of a chemical potential $\mu$ and a temperature $T_i$ via a classical Metropolis algorithm. This corresponds to the initial cloud of the experiment.

	To imitate coherent splitting of the initial cloud into two clouds, we consider a second state $\psi_2$ and initialize it with quantum fluctuations \cite{Mathey2017}.  
	We then use a $\pi/2$-pulse rotation as a quench to initialize non-equilibrium states $\psi_1$ and $\psi_2$ with equal densities, in a similar manner to the method employed in Ref.\ \cite{Mathey2017}. 
	Following the quench, $\psi_1$ and $\psi_2$  evolve under the equations of motion
	\begin{align} 
	i\hbar \frac{\partial \psi_1}{\partial  t} &= \Bigl(- \frac{\hbar^2}{2m} \nabla^2 + V(r) + g_{\mathrm{2D}}  |\psi_1|^2 \Bigr) \psi_1 + J_t \psi_2   ,  \label{eq_eom1} \\
	i\hbar \frac{\partial \psi_2}{\partial  t} &= \Bigl(- \frac{\hbar^2}{2m} \nabla^2 + V(r) +  g_{\mathrm{2D}} |\psi_2|^2  \Bigr) \psi_2   + J_t \psi_1   ,  \label{eq_eom2}
	\end{align}
	where we have added a time-dependent tunneling term $J_t= J\exp(-t/t_0)$ to account for nonzero coupling of the clouds during and shortly after the splitting.  $J= \hbar^2/(2m l^2)$ is the single-particle tunneling energy. 
	We set $t_0=60\, \mms$ to capture the rapid decoupling dynamics of the experiment after the splitting.
	We use $\omega_r/(2\pi) = 11\,  \mHz$, which is the same as experiment, and calculate the time evolution of $\psi_1(t)$ and $\psi_2(t)$ to analyze the dynamics after the quench at $t=0$. 
	From the arguments of complex numbers $\psi_1(t)$ and $\psi_2(t)$, we calculate the relative-phase correlation function and the vortex density, in the same way as the experiment, and average over the initial ensemble. 
	The initial temperature $T_i$ is chosen to be close to the experimental temperature. 
	In Fig.\ \ref{fig:simScaling}, we plot the time evolution of $\eta$ and $n_v$ against the rescaled time $t/t_c$ where $t_c$ is the temperature-dependent crossover time, obtained from the simulation data.
	%At long times $t/t_c \gtrsim 1$ and for higher temperature, 


\begin{thebibliography}{49}
\bibitem [{sca()}]{scaling}%
		\BibitemOpen
		\href@noop {} {}\bibinfo {note} {To motivate the scaling regimes, we consider
			a linear spectrum $\omega_k = c k$. With this dispersion, dynamical phase
			factors of the form $\exp(i \omega_k t)$ are kept invariant, if a $z=1$
			scaling is applied that leaves $c k t$ invariant. Furthermore, for a spectrum
			that includes an additional length $\xi$, such as $\omega_k = c \sqrt{k^2 +
				1/\xi^2(t)}$, for long times, only the modes with $\omega \approx c/\xi(t)$
			contribute to the dynamics. Here the scaling is replaced by keeping $c
			t/\xi(t)$ fixed, as it was demonstrated for the long-time dynamics in Fig.\
			2C. As discussed in the main text and shown in Fig.\ 4, the length scale
			$\xi(t)$ is related to the vortex fugacity increasing at long
			times.}\BibitemShut {Stop}%
\end{thebibliography}
\end{document}